\documentclass[aps,prd,
noeprint,
preprint,
amsmath,
amssymb,
superscriptaddress
]{revtex4-1}

\usepackage[utf8]{inputenc}
\usepackage{graphicx}
\usepackage{dcolumn}
\usepackage{bm}
\usepackage{hyperref}
\usepackage{upgreek,textgreek}
\usepackage{xcolor}
\usepackage{longtable}

\begin{document}

\title{QCD Corrections to $e^+e^- \rightarrow H^{\pm}W^{\mp}$ in Type-I THDM at Electron Positron Colliders}

\author{Qiang Yang}
\affiliation{State Key Laboratory of Particle Detection and Electronics, University of Science and Technology of China, Hefei 230026, Anhui, People's Republic of China}
\affiliation{Department of Modern Physics, University of Science and Technology of China, Hefei 230026, Anhui, People's Republic of China}

\author{Ren-You Zhang}
\email{zhangry@ustc.edu.cn}
\affiliation{State Key Laboratory of Particle Detection and Electronics, University of Science and Technology of China, Hefei 230026, Anhui, People's Republic of China}
\affiliation{Department of Modern Physics, University of Science and Technology of China, Hefei 230026, Anhui, People's Republic of China}

\author{Ming-Ming Long}
\affiliation{State Key Laboratory of Particle Detection and Electronics, University of Science and Technology of China, Hefei 230026, Anhui, People's Republic of China}
\affiliation{Department of Modern Physics, University of Science and Technology of China, Hefei 230026, Anhui, People's Republic of China}

\author{Shao-Ming Wang}
\affiliation{Department of Physics, Chongqing University, Chongqing 401331, People's Republic of China}

\author{Wen-Gan Ma}
\affiliation{State Key Laboratory of Particle Detection and Electronics, University of Science and Technology of China, Hefei 230026, Anhui, People's Republic of China}
\affiliation{Department of Modern Physics, University of Science and Technology of China, Hefei 230026, Anhui, People's Republic of China}

\author{Jian-Wen Zhu}
\affiliation{State Key Laboratory of Particle Detection and Electronics, University of Science and Technology of China, Hefei 230026, Anhui, People's Republic of China}
\affiliation{Department of Modern Physics, University of Science and Technology of China, Hefei 230026, Anhui, People's Republic of China}

\author{Yi Jiang}
\affiliation{State Key Laboratory of Particle Detection and Electronics, University of Science and Technology of China, Hefei 230026, Anhui, People's Republic of China}
\affiliation{Department of Modern Physics, University of Science and Technology of China, Hefei 230026, Anhui, People's Republic of China}

\date{\today}

\begin{abstract}
We investigate in detail the charged Higgs production associated with a $W$ boson at electron-positron colliders within the framework of the Type-I two-Higgs-doublet model (THDM). We calculate the integrated cross section at the LO and analyze the dependence of the cross section on the THDM parameters and the colliding energy in a benchmark scenario of the input parameters of Higgs sector. The numerical results show that the integrated cross section is sensitive to the charged Higgs mass, especially in the vicinity of $m_{H^{\pm}} \simeq 184~ {\rm GeV}$ at a $500~ {\rm GeV}$ $e^+e^-$ collider, and decreases consistently as the increment of $\tan\beta$ in the low $\tan\beta$ region. The peak in the colliding energy distribution of the cross section arises from the resonance of loop integrals and its position moves towards low colliding energy as the increment of $m_{H^{\pm}}$. We also study the two-loop NLO QCD corrections to both the integrated cross section and the angular distribution of the charged Higgs boson, and find that the QCD relative correction is also sensitive to the charged Higgs mass and strongly depends on the final-state phase space. For $\tan\beta = 2$, the QCD relative correction at a $500~ {\rm GeV}$ $e^+e^-$ collider varies in the range of $[-10\%,\, 11\%]$ as $m_{H^{\pm}}$ increases from $150$ to $400~ {\rm GeV}$.

\begin{description}
\item[keywords]
two-loop QCD correction, Type-I THDM, charged Higgs boson
\end{description}
\end{abstract}

\maketitle

\section{\label{sec:introduction}Introduction}
\par
In July 2012, the $125~ {\rm GeV}$ Higgs boson has been discovered by ATLAS and CMS collaborations at CERN Large Hadron Collider (LHC) \cite{Aad:2012tfa,Chatrchyan:2012xdj}. In addition to measuring the $125~ {\rm GeV}$ Higgs boson precisely, great efforts have been made to search for exotic Higgs bosons in various scenarios beyond the standard model (SM). Among all the extensions of the SM, the two-Higgs-doublet model (THDM) is an appealing one. It provides rich phenomena such as charged Higgs bosons, explicit and spontaneous $\mathcal{CP}$-violation, and the candidate for dark matter, since the Higgs sector of the THDM is composed of two complex scalar doublets \cite{Lee:1973iz,Gunion:1989we,Gunion:2002zf,Branco:2011iw}. In the THDM, there are five Higgs bosons: two neutral $\mathcal{CP}$-even Higgs bosons $h$ and $H$ ($m_h < m_H$), two charged Higgs bosons $H^{\pm}$ and a neutral $\mathcal{CP}$-odd Higgs boson $A$. Although both of the two $\mathcal{CP}$-even scalars can be interpreted as the $125~ {\rm GeV}$ Higgs boson in the alignment limit, we assume that the lighter scalar $h$ is the $125~ {\rm GeV}$ Higgs boson in this paper. Since the flavor changing neutral currents (FCNCs) can be induced in THDM which have not been observed, an additional $Z_2$ symmetry is imposed to eliminate FCNCs at the tree level. Depending on the types of the Yukawa interactions between fermions and the two Higgs doublets, one can introduce several different types of THDMs (Type-I, Type-II, lepton-specific, and flipped) \cite{Branco:2011iw}. The most investigated THDMs are the Type-I and Type-II THDMs. In the Type-I THDM, all the fermions only couple to one of the two Higgs doublets, while in the Type-II THDM, the up-type and down-type fermions couple to the two Higgs doublets, respectively.

\par
The THDMs have been widely studied in many different aspects in previous works. Since it is possible to introduce the spontaneous $\mathcal{CP}$-violation in THDM, it has been considered as a solution to the problem of baryogenesis in Refs.\cite{Turok:1990zg,Davies:1994id,Fromme:2006cm}. In Refs.\cite{Deshpande:1977rw,Dolle:2009fn}, the neutral scalar in the inert THDM is interpreted as the candidate for dark matter. The existence of charged Higgs boson is one important characteristic of new physics beyond the SM. Therefore, searching for charged Higgs boson in various aspects is a high priority of experiments. At the LHC Run II, the charged Higgs boson has been probed in various channels, such as $H^{\pm} \rightarrow tb,\, \tau \nu_\tau~ \text{and}~ WZ$ \cite{Sirunyan:2017sbn,Sirunyan:2019arl,Aaboud:2018gjj,Aaboud:2018cwk}.

\par
To match the precise experimental data, the theoretical predictions on kinematic observables should be calculated with high precision. The renormalization of the THDM has been detailedly studied in different renormalization schemes in Refs.\cite{Santos:1996vt,Degrande:2014vpa,Krause:2016oke,Denner:2016etu,Altenkamp:2017ldc}. The production mechanisms and decay modes of the charged Higgs boson have been investigated at one-loop level in the THDM. The Drell-Yan production of charged Higgs pair has been studied at NLO in Refs.\cite{Arhrib:1998gr,Heinemeyer:2016wey}. The full one-loop contributions for the charged Higgs production associated with a vector boson were given in Refs.\cite{Zhu:1999ke,Arhrib:1999rg,Kanemura:1999tg,Heinemeyer:2016wey}. The dominant decays of the charged Higgs boson into $tb$ and $\tau \nu_{\tau}$ have been studied in Refs.\cite{Drees:1991nf,Roy:1999xw}, and the loop-induced decay modes $H^{\pm}\to W^{\pm}\gamma$ and $H^{\pm}\to W^{\pm}Z$ have also been investigated in Refs.\cite{CapdequiPeyranere:1990qk,Kanemura:1997ej,HernandezSanchez:2004tq,Arhrib:2006wd}. In this work, we focus on the $H^{\pm}W^{\mp}$ associated production at electron-positron colliders in the Type-I THDM. This production channel is a loop-induced process at the lowest order due to the absence of tree-level $H^{\pm}W^{\mp}\gamma$ and $H^{\pm}W^{\mp}Z$ couplings, and has been investigated at one-loop level in the THDM as well as the minimal supersymmetric standard model \cite{Farris:2003qj,Kanemura:2011kc,Logan:2002jh,Logan:2002it}. In order to test the THDM via $H^{\pm}W^{\mp}V$ couplings precisely, we study in detail the two-loop QCD corrections to the $e^+e^- \rightarrow H^{\pm}W^{\mp}$ process, provide the NLO QCD corrected integrated and differential cross sections, and discuss the dependence on the THDM parameters and the $e^+e^-$ colliding energy.

\par
The rest of this paper is organized as follows. In Sec.\ref{sec:THDM}, we give a brief review of the Type-I THDM and provide the benchmark scenario that we adopt. The methods and details of our LO and NLO calculations are presented in Sec.\ref{sec:calculation}. In Sec.\ref{sec:results}, the numerical results for both integrated and differential cross sections and some discussions are provided. Finally, a short summary is given in Sec.\ref{sec:summary}.

\section{\label{sec:THDM}Two-Higgs-Doublet model}
\par
The Higgs sector of the THDM is composed of two complex scalar doublets $\Phi_1=(\phi_{1}^{+},\phi_{1}^0)^{T}$ and $\Phi_2=(\phi_{2}^{+},\phi_{2}^0)^{T}$, which are both in the $(\mathbf{1}, \mathbf{2}, \mathbf{1})$ representation of the $SU(3)_C \otimes SU(2)_L \otimes U(1)_Y$ gauge group. In this paper, we consider only the $\mathcal{CP}$-conserving THDM with a discrete $Z_2$ symmetry of the form $\Phi_1 \rightarrow -\Phi_1$. Then the renormalizable and gauge invariant scalar potential is given by
\begin{widetext}
\begin{equation}
\label{eq:potential}
\begin{aligned}
\mathcal{V}_{\text {scalar}}
& =
m_{11}^{2} \Phi_{1}^{\dagger} \Phi_{1}
+
m_{22}^{2} \Phi_{2}^{\dagger} \Phi_{2}
-
\left[ m_{12}^{2} \Phi_{1}^{\dagger} \Phi_{2} + \text{h.c.} \right]
+
\frac{1}{2} \lambda_{1} \left( \Phi_{1}^{\dagger} \Phi_{1} \right)^{2}
+
\frac{1}{2} \lambda_{2} \left( \Phi_{2}^{\dagger} \Phi_{2} \right)^{2}
\\
& +
\lambda_{3} \left( \Phi_{1}^{\dagger} \Phi_{1} \right) \left( \Phi_{2}^{\dagger} \Phi_{2} \right)
+
\lambda_{4} \left( \Phi_{1}^{\dagger} \Phi_{2} \right) \left( \Phi_{2}^{\dagger} \Phi_{1} \right)
+
\frac{1}{2} \left[ \lambda_{5} \left( \Phi_{1}^{\dagger} \Phi_{2} \right)^{2} + \text{h.c.} \right].
\end{aligned}
\end{equation}
\end{widetext}
Since the parameter $m_{12}$ has the mass-dimension $1$, the terms of this kind only break the $Z_2$ symmetry softly which can be retained. The parameters $m_{11},\, m_{22},\, \lambda_{1},\, \lambda_{2},\, \lambda_{3},\, \lambda_{4}$ have to be real since the Lagrangian must be real. Though the parameters $m_{12}$ and $\lambda_{5}$ can be complex, the imaginary parts of these two parameters would induce explicit $\mathcal{CP}$ violation that we do not consider in this paper. So we assume all the parameters in Eq.(\ref{eq:potential}) are real. The minimization of the potential in Eq.(\ref{eq:potential}) gives two minima $\langle\Phi_{1}\rangle$ and $\langle\Phi_{2}\rangle$ of the form
\begin{equation}
    \left\langle\Phi_{1}\right\rangle=
    \left(\begin{array}{c}{0} \\ {v_{1}/\sqrt{2} }\end{array}\right),
    \quad
    \left\langle\Phi_{2}\right\rangle=
    \left(\begin{array}{c}{0} \\ {v_{2}/\sqrt{2} }\end{array}\right),
\end{equation}
where $v_{1}$ and $v_{2}$ are the vacuum expectation values of the neutral components of the two Higgs doublets $\Phi_{1}$ and $\Phi_{2}$, respectively. With respect to the convention in Ref.\cite{Eriksson:2010zzb}, we define $v_1= v\cos\beta$ and $v_2 = v\sin\beta$, where $v=(\sqrt{2}G_{F})^{-1/2}\approx 246~ {\rm GeV}$. Expanding at the minima, the two complex Higgs doublets $\Phi_{1,2}$ can be expressed as
\begin{equation}
    \Phi_{1}=
    \left(\begin{array}{c}{\phi_{1}^{+}} \\ {(v_{1}+\rho_{1}+i\eta_{1})/\sqrt{2} }\end{array}\right),
    \quad
    \Phi_{2}=
    \left(\begin{array}{c}{\phi_{2}^{+}} \\ {(v_{2}+\rho_{2}+i\eta_{2})/\sqrt{2} }\end{array}\right).
\end{equation}
The mass eigenstates of the Higgs fields are given by \cite{Branco:2011iw,Altenkamp:2017ldc}
\begin{equation}
    \left(\begin{array}{c}{H} \\ {h}\end{array}\right)=
    \left(\begin{array}{cc}{\cos \alpha} & {\sin \alpha} \\ {-\sin \alpha} & {\cos \alpha}\end{array}\right)\left(\begin{array}{c}{\rho_{1}} \\ {\rho_{2}}\end{array}\right),
\end{equation}
\begin{equation}
    \left(\begin{array}{c}{G^{0}} \\ {A}\end{array}\right)=
    \left(\begin{array}{cc}{\cos \beta} & {\sin \beta} \\ {-\sin \beta} & {\cos \beta}\end{array}\right)\left(\begin{array}{c}{\eta_{1}} \\ {\eta_{2}}\end{array}\right),
\end{equation}
\begin{equation}
    \left(\begin{array}{c}{G^{\pm}} \\ {H^{\pm}}\end{array}\right)=
    \left(\begin{array}{cc}{\cos \beta} & {\sin \beta} \\ {-\sin \beta} & {\cos \beta}\end{array}\right)\left(\begin{array}{c}{\phi_{1}^{\pm}} \\ {\phi_{2}^{\pm}}\end{array}\right),
\end{equation}
where $\alpha$ is the mixing angle of the neutral $\mathcal{CP}$-even Higgs sector. After the spontaneous electroweak symmetry breaking, the charged and neutral Goldstone fields $G^{\pm}$ and $G^{0}$ are absorbed by the weak gauge bosons $W^{\pm}$ and $Z$, respectively. Thus, the THDM predicts the existence of five physical Higgs bosons: two neutral $\mathcal{CP}$-even Higgs bosons $h$ and $H$, one neutral $\mathcal{CP}$-odd Higgs boson $A$, and two charged Higgs bosons $H^{\pm}$. The masses of these physical Higgs bosons are given by 
\begin{equation}
    \begin{aligned}
        m_{A}^{2}       & = \frac{m_{12}^{2}}{\sin \beta \cos \beta}-v^{2} \lambda_{5},                                                                                                                    \\
        m_{H^{\pm}}^{2} & = \frac{m_{12}^{2}}{\sin \beta \cos \beta}-\frac{v^{2}}{2}\left(\lambda_{4}+\lambda_{5}\right),                                                                                  \\
        m_{H, h}^{2}    & =\frac{1}{2}\left[\mathcal{M}_{11}^{2}+\mathcal{M}_{22}^{2} \pm \sqrt{\left(\mathcal{M}_{11}^{2}-\mathcal{M}_{22}^{2}\right)^{2}+4\left(\mathcal{M}_{12}^{2}\right)^{2}}\right],
    \end{aligned}
\end{equation}
where $\mathcal{M}^2$ is the mass matrix of the neutral $\mathcal{CP}$-even Higgs sector. The explicit form of $\mathcal{M}^2$ is expressed as
\begin{equation}
    \mathcal{M}^{2}=
    m_{A}^{2}\left(\begin{array}{cc}
            {\sin^{2} \beta}         & {-\sin \beta \cos \beta} \\
            {-\sin \beta \cos \beta} & {\cos^{2}\beta}\end{array}\right)
    +v^{2} \mathcal{B}^{2},
\end{equation}
where
\begin{equation}
    \mathcal{B}^{2}=
    \left(\begin{array}{cc}
            {\lambda_{1} \cos^{2}\beta + \lambda_{5} \sin^{2}\beta}    & {\left(\lambda_{3}+\lambda_{4}\right) \sin\beta \cos\beta} \\
            {\left(\lambda_{3}+\lambda_{4}\right) \sin\beta \cos\beta} & {\lambda_{2} \sin^{2}\beta+\lambda_{5} \cos^{2}\beta}
        \end{array}\right).
\end{equation}
The SM Higgs is the combination of the two neutral $\mathcal{CP}$-even Higgs bosons as
\begin{equation}
    \begin{aligned}
        h^{SM} & =  \rho_{1} \cos\beta + \rho_{2} \sin\beta           \\
               & =         h\sin(\beta-\alpha) + H\cos(\beta-\alpha).
    \end{aligned}
\end{equation}
Thus, the lighter neutral $\mathcal{CP}$-even scalar $h$ can be identified as the SM-like Higgs boson in the so-called alignment limit of $\sin(\beta-\alpha)\rightarrow 1$. In this paper, we consider the lighter $\mathcal{CP}$-even Higgs $h$ as the SM-like Higgs boson discovered at the LHC.

The input parameters for the Higgs sector of the THDM are chosen as
\begin{equation}
    \left\{ m_{h},\, m_{H},\, m_{A},\, m_{H^{\pm}},\, m_{12},\, \sin(\beta-\alpha),\, \tan\beta \right\},
\end{equation}
which are implemented as the ``physical basis'' in \textit{2HDMC} \cite{Eriksson:2010zzb}. We adopt the following benchmark scenario,
\begin{equation}
    \begin{aligned}
         & m_{h}=125.18 ~{\rm GeV},                 \\
         & m_{H} = m_{A} = m_{H^{\pm}},              \\
         & m_{12}^{2} = m_{A}^{2}\sin\beta\cos\beta, \\
         & \sin(\beta-\alpha) = 1,                   \\
         & m_{H^{\pm}} \in [150, 400] ~{\rm GeV},    \\
         & \tan\beta \in [1, 5],
    \end{aligned}
    \label{eq:benchmark}
\end{equation}
which satisfies the theoretical constraints from perturbative unitarity \cite{Grinstein:2015rtl}, stability of vacuum \cite{Nie:1998yn}, and tree-level unitarity \cite{Akeroyd:2000wc}. The $Z_{2}$ soft-breaking parameter $m_{12}^2$ is chosen as $m_{A}^{2}\sin\beta\cos\beta$ in order to satisfy the perturbative unitarity for $\tan\beta \in [1, 5]$. Considering the constraints from experiments at the $13~ {\rm TeV}$ LHC in Ref.\cite{Sirunyan:2019xls}, $\cos(\beta-\alpha)$ is very closed to $0$. So we apply the alignment limit in our benchmark scenario to set $\sin(\beta-\alpha) = 1$.
As to the $m_{H^{\pm}}$ and $\tan\beta$ parameters, we concentrate on the region with low mass and small $\tan\beta$.

\section{\label{sec:calculation}Descriptions of calculation}
\par
In this section, we present in detail the calculation procedure for the $e^{+}e^{-}\rightarrow H^{\pm}W^{\mp}$ process at one- and two-loop levels. The Feynman diagrams and the amplitudes are generated by \textit{FeynArts-3.11} \cite{Hahn:2000kx}, using the Feynman rules of THDM in Ref.\cite{Altenkamp:2017ldc}. The evaluation of Dirac trace and the contraction of Lorentz indices are performed by the \textit{FeynCalc-9.3} package \cite{Mertig:1990an,Shtabovenko:2016sxi}. In order to reduce the Feynman integrals into the combinations of a small set of integrals called the master integrals (MIs), we utilize the \textit{KIRA-1.2} package \cite{Maierhoefer:2017hyi}, which adopts the integration-by-parts (IBP) method with Laporta's algorithm \cite{Laporta:2001dd}. One can get the numerical results of amplitudes after evaluating the MIs which is the main difficulty of multi-loop calculation.

\par
In this paper, we calculate the MIs by using the ordinary differential equations (ODEs) method \cite{Liu:2017jxz}.
A $L$-loop Feynman integral can be expressed as
\begin{equation}
    \mathcal{I}(\left\{a_{1}, \ldots, a_{n} \right\},\, D,\, \eta) = \int \prod_{j=1}^{L}d^{D}l_{j}\prod_{k=1}^{n}\frac{1}{(E_{k}+i\eta)^{a_{k}}},
\end{equation}
where $D\equiv 4-2\epsilon$ and $E_{k} = q_{k}^{2}-m_{k}^2$ are the denominators of Feynman propagators in which the $q_{k}$ are the linear combinations of loop momenta and external momenta. The physical results of the Feynman integrals are obtained by taking $\eta \rightarrow 0^+$, i.e.,
\begin{equation}
    \mathcal{I}(\left\{a_{1}, \ldots, a_{n} \right\},\, D,\, 0) = \lim_{\eta\to 0^{+}} \mathcal{I}(\left\{a_{1}, \ldots, a_{n} \right\},\, D,\, \eta).
\end{equation}
One can construct the ODEs with respect to $\eta$,
\begin{equation}
    \frac{\partial\vec{\mathcal{I}}(\eta)}{\partial\eta} = \mathcal{M}(\eta).\vec{\mathcal{I}}(\eta),
\end{equation}
where $\vec{\mathcal{I}}$ is a complete set of MIs. The boundary conditions of these ODEs are chosen at $\eta = \infty$ which are the simple vacuum integrals. The analytical expressions for the vacuum integrals up to three-loop order can be found in Refs.\cite{Davydychev:1992mt,Broadhurst:1998rz,Kniehl:2017ikj}. To solve these ODEs numerically, we utilize the \textit{odeint} package \cite{Ahnert:2011} to evaluate ODEs from an initial point $\eta_i$ to a target point $\eta_j$. To perform the asymptotic expansion in the domain nearby $\eta=0$, we transform the coefficient matrix into normalized fuchsian form with the help of the \textit{epsilon} package \cite{Prausa:2017ltv}.

\subsection{\label{subsec:LO} Calculation at LO}
\par
\begin{figure*}
    \centering
    \includegraphics[scale=0.4]{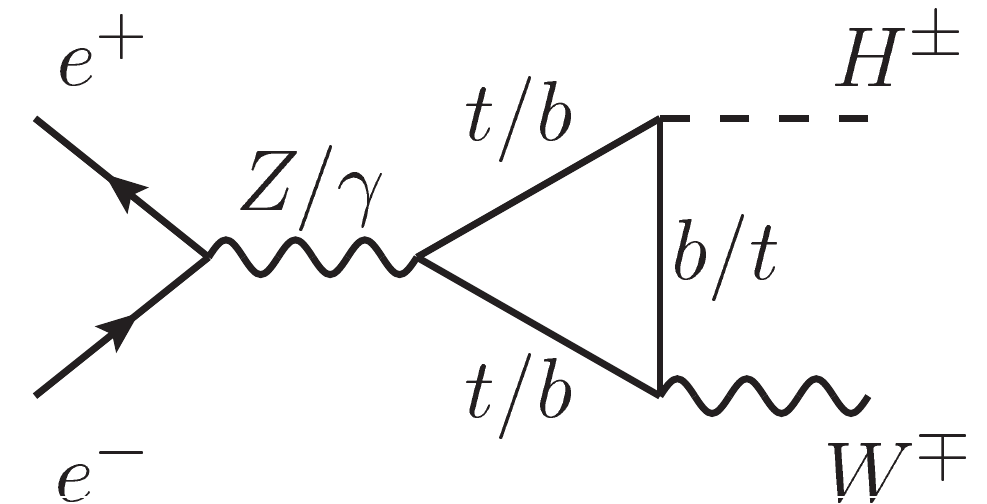}
    \includegraphics[scale=0.4]{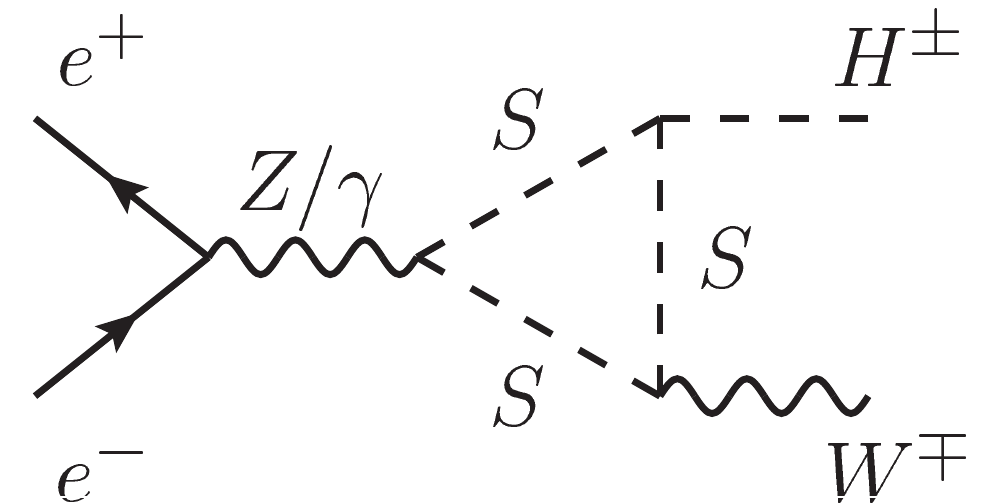}
    \includegraphics[scale=0.4]{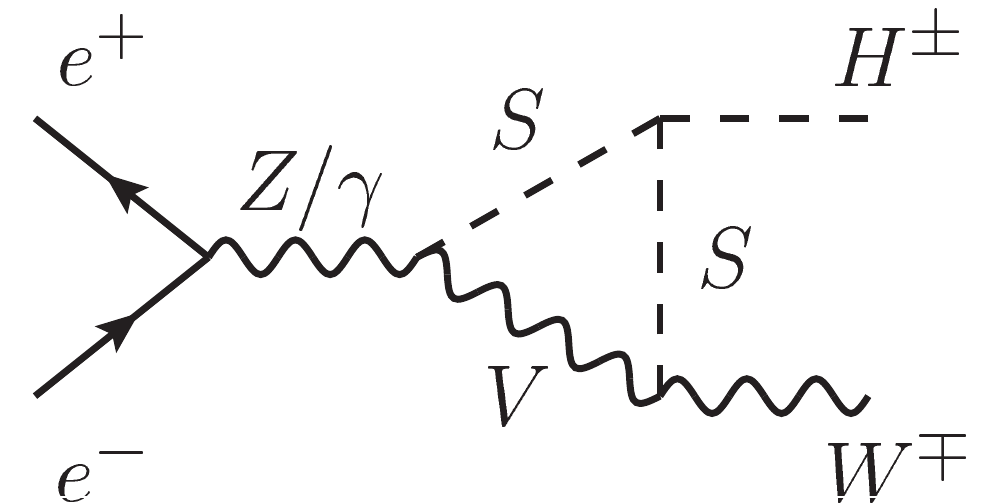}
    \includegraphics[scale=0.4]{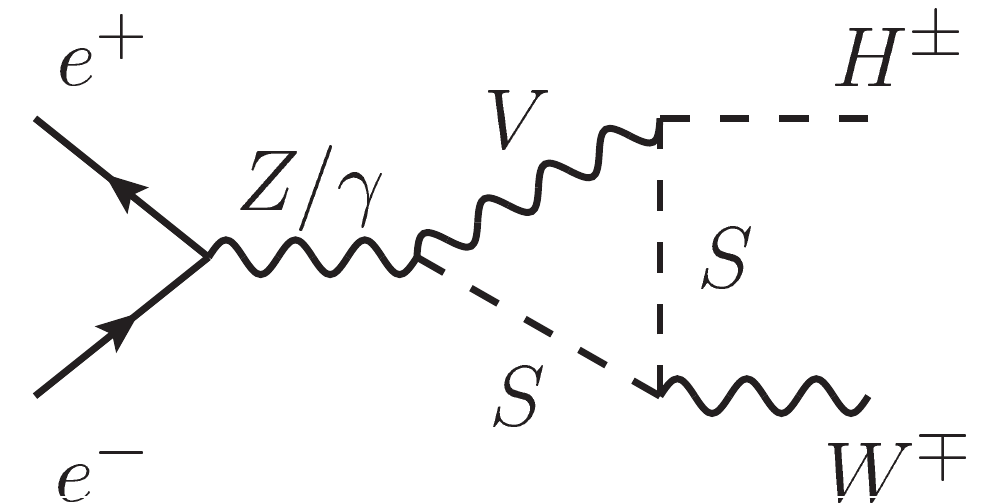}
    \includegraphics[scale=0.4]{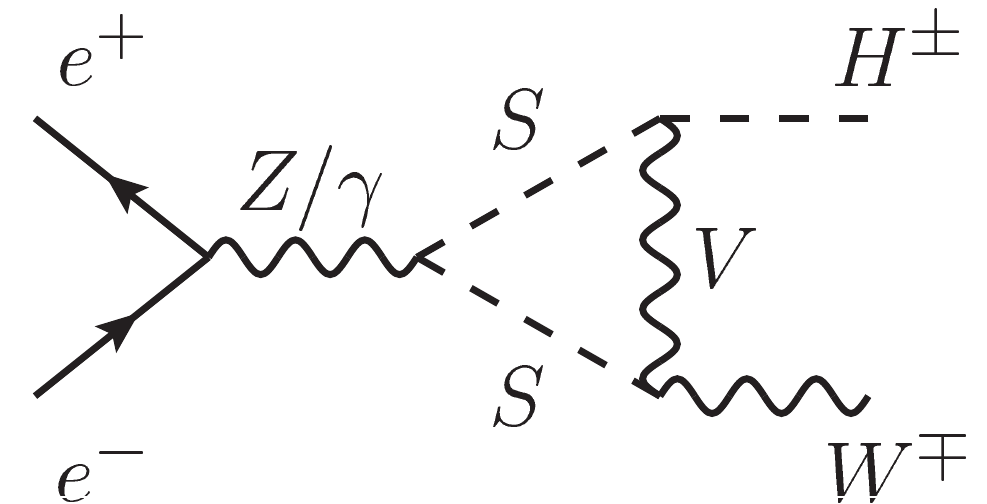}
    \includegraphics[scale=0.4]{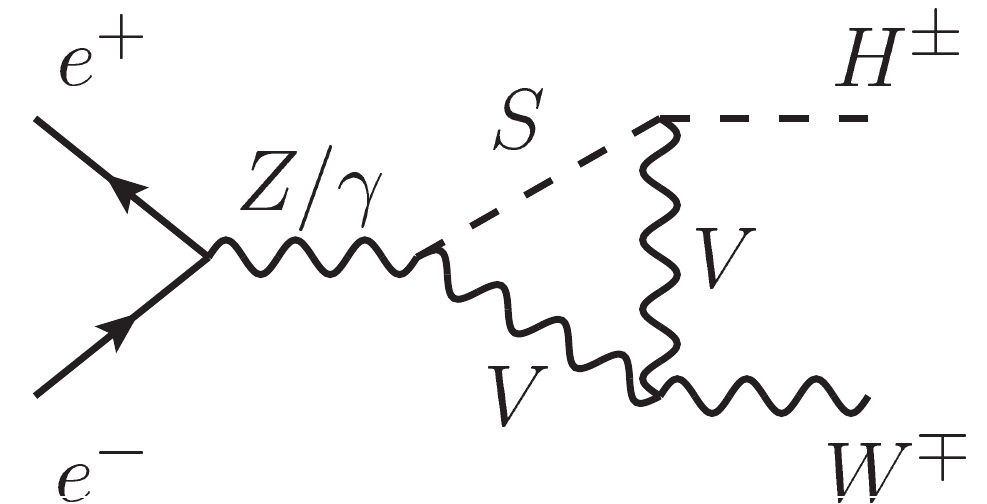}
    \includegraphics[scale=0.4]{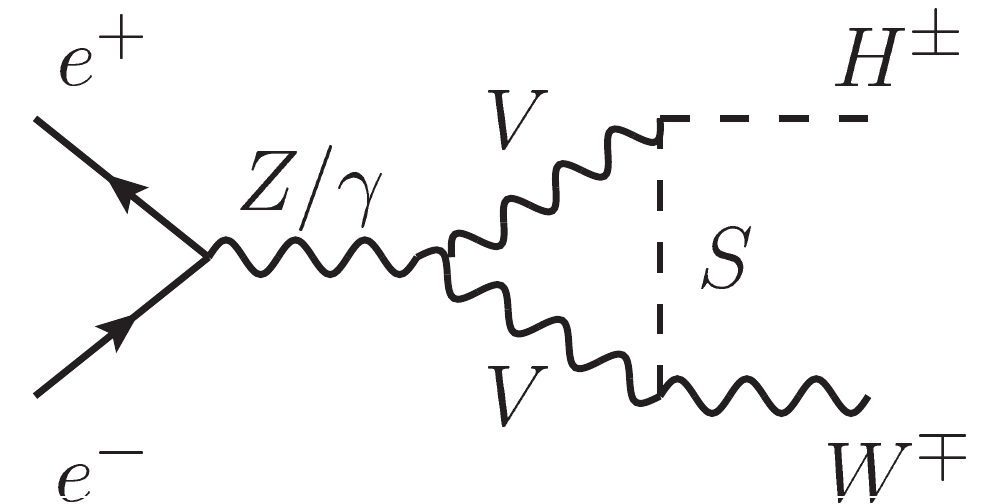}
    \includegraphics[scale=0.4]{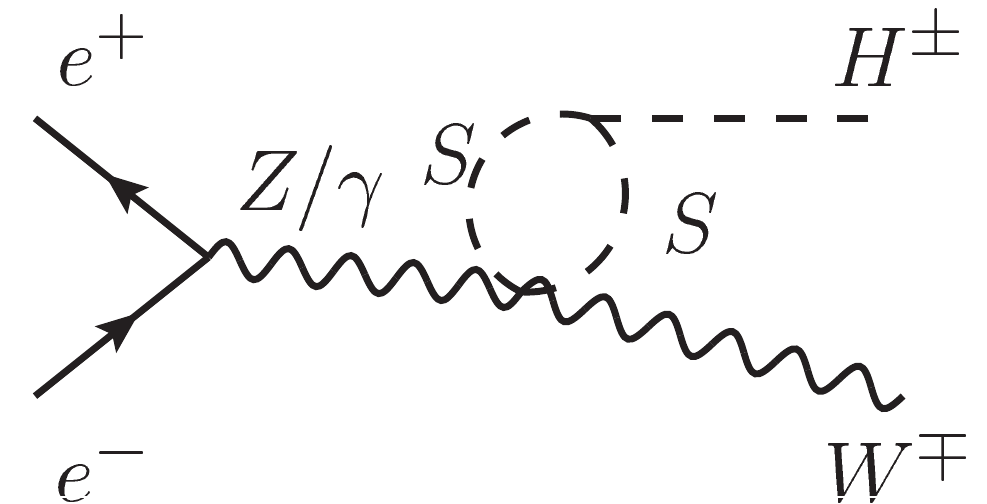}
    \includegraphics[scale=0.4]{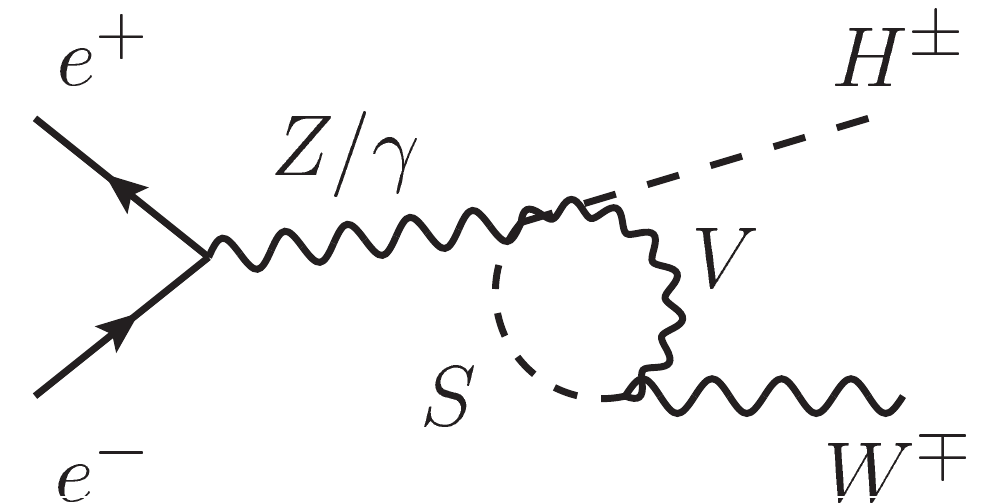}
    \includegraphics[scale=0.4]{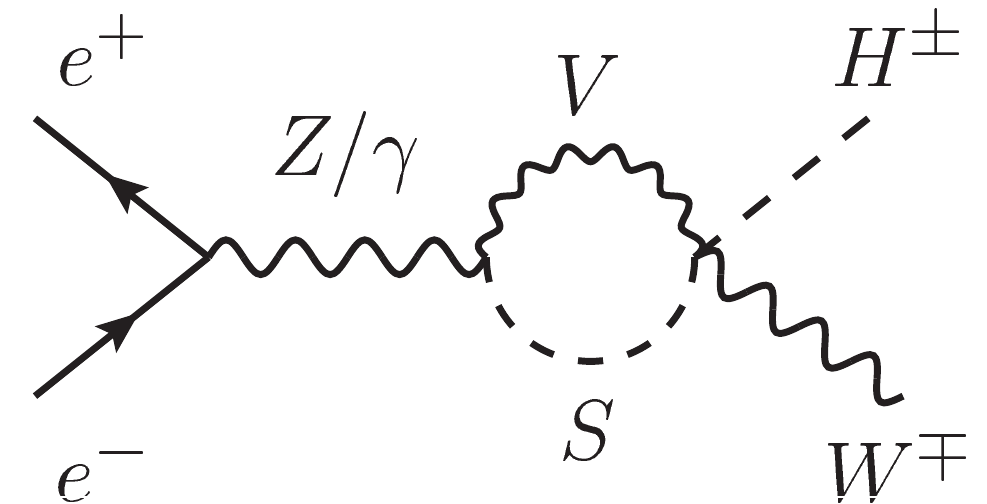}
    \includegraphics[scale=0.4]{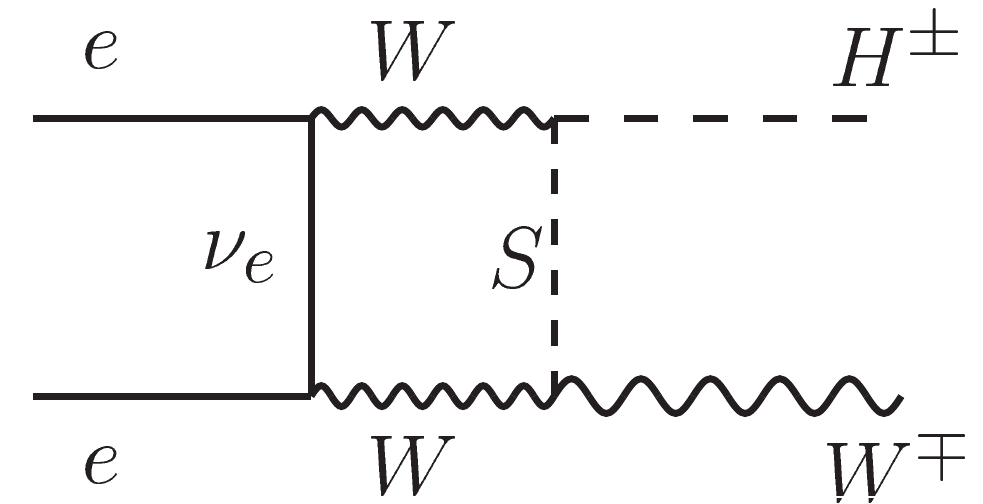}
    \includegraphics[scale=0.4]{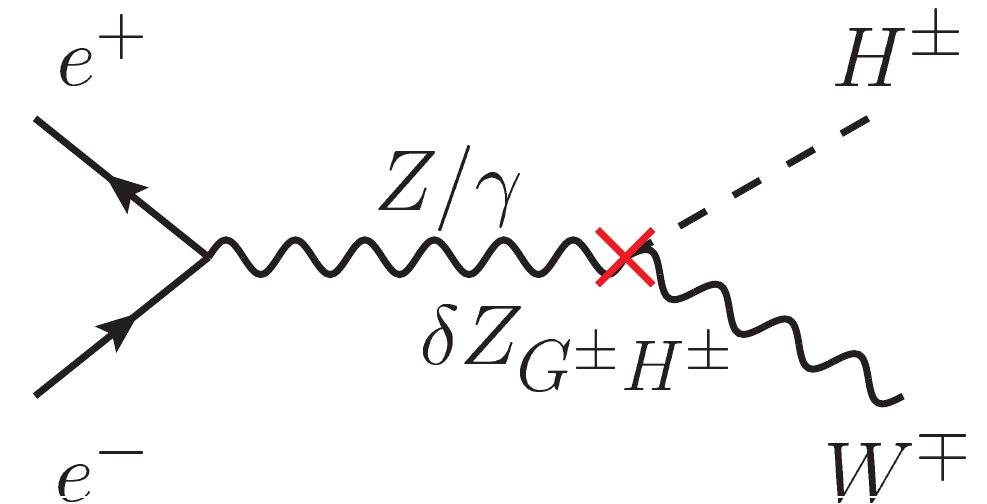}
    \caption{Representative Feynman diagrams for $e^{+}e^{-} \rightarrow H^{\pm}W^{\mp}$ at LO.}
    \label{fig:feynman_lo}
\end{figure*}

For the LO calculation, we adopt the 't Hooft-Feynman gauge with the on-shell renormalization scheme at one-loop order mentioned in Ref.\cite{Altenkamp:2017ldc}. Some representative Feynman diagrams for the $e^{+}e^{-} \rightarrow H^{\pm}W^{\mp}$ process at the LO are shown in Fig.\ref{fig:feynman_lo}, where $S$ and $V$ in the loops represent the Higgs/Goldstone and weak gauge bosons, respectively. Due to the tiny mass of electron, the contribution from the diagrams involving Higgs Yukawa coupling to electron is ignored. The diagrams with $V-S$ mixing are also not shown in Fig.\ref{fig:feynman_lo}, because these diagrams can induce a factor of $m_e$ via Dirac equation. The last diagram in Fig.\ref{fig:feynman_lo} is a vertex counterterm diagram induced by the renormalization constant $\delta Z_{G^{\pm}H^{\pm}}$ at one-loop level. In the on-shell renormalization scheme, the renormalization constant $\delta Z_{G^{\pm}H^{\pm}}$ is given by
\begin{equation}
    \delta Z_{G^{\pm}H^{\pm}} = - \frac{2 \widetilde{Re} \sum^{W^{\pm}H^{\pm}}(m_{H^{\pm}}^2)}{m_{W}},
\label{eq:renconst}
\end{equation}
where $\sum^{W^{\pm}H^{\pm}}(m_{H^{\pm}}^2)$ is the transition of $W^{\pm}-H^{\pm}$ at $p^{2} = m_{H^{\pm}}^2$, and $\widetilde{Re}$ means to take the real parts of the loop integrals in the transition. It is worth mentioning that Eq.(\ref{eq:renconst}) is valid at both $\mathcal{O}(\alpha)$ and $\mathcal{O}(\alpha\alpha_s)$.

\subsection{\label{subsec:NLO} Calculation at QCD NLO}
\par
\begin{figure*}
    \centering
    \includegraphics[scale=0.4]{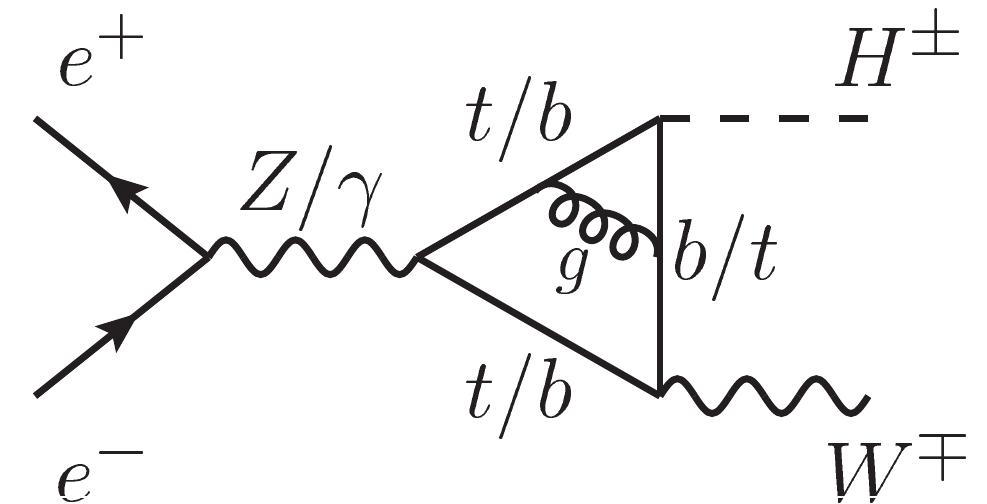}
    \includegraphics[scale=0.4]{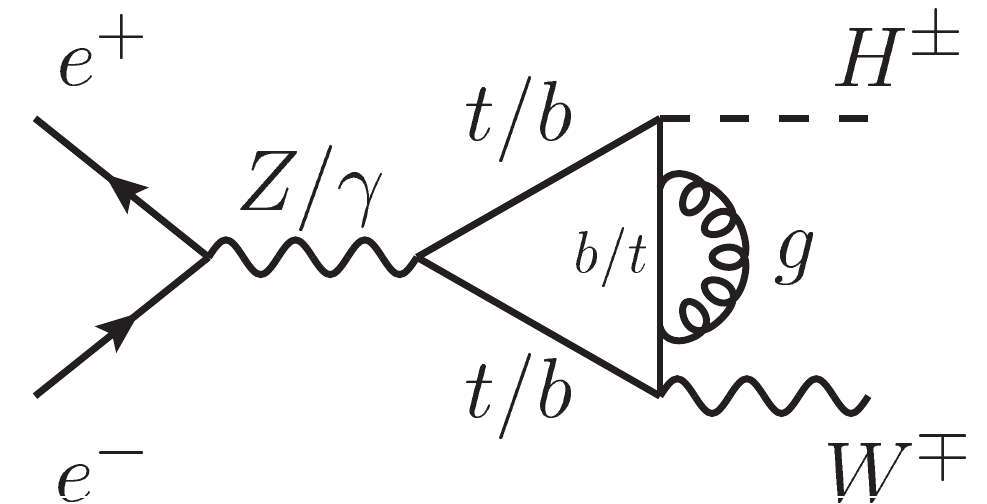}
    \includegraphics[scale=0.4]{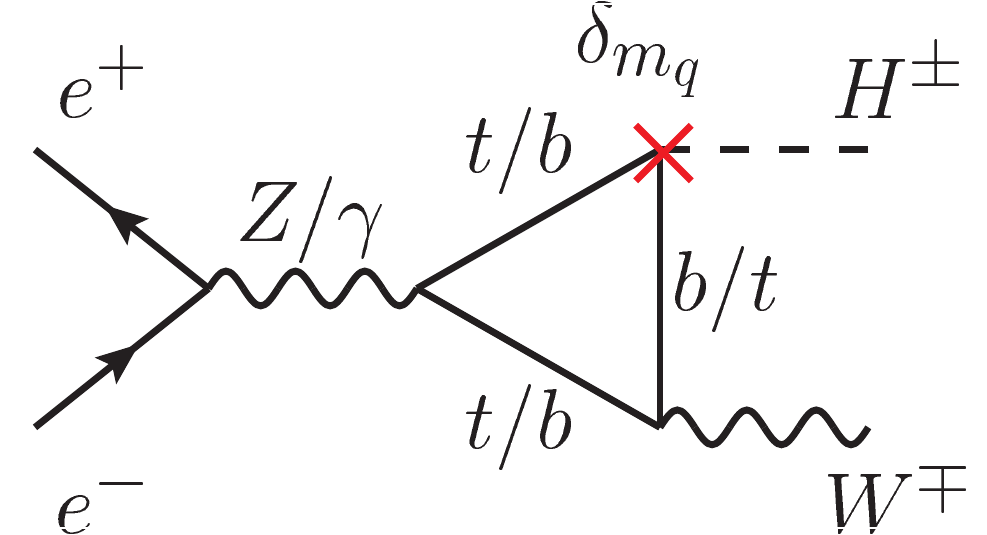}
    \includegraphics[scale=0.4]{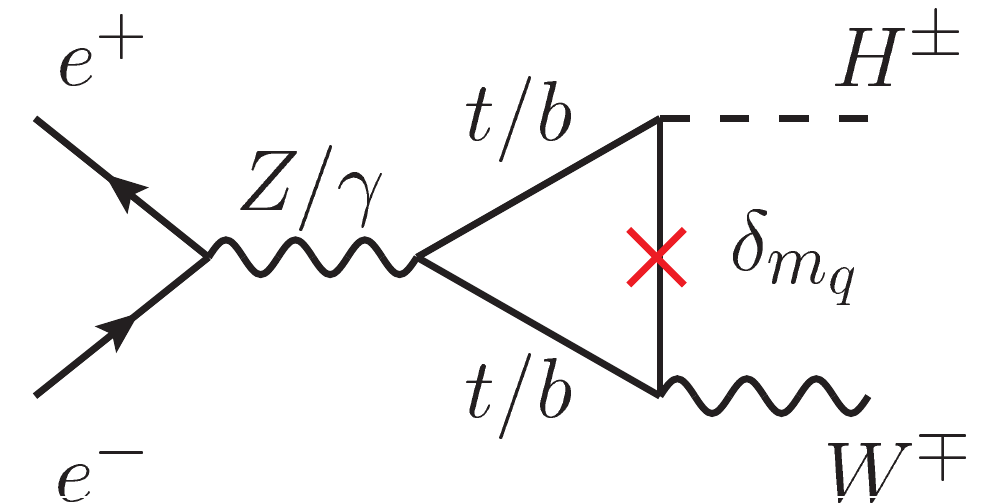}
    \includegraphics[scale=0.4]{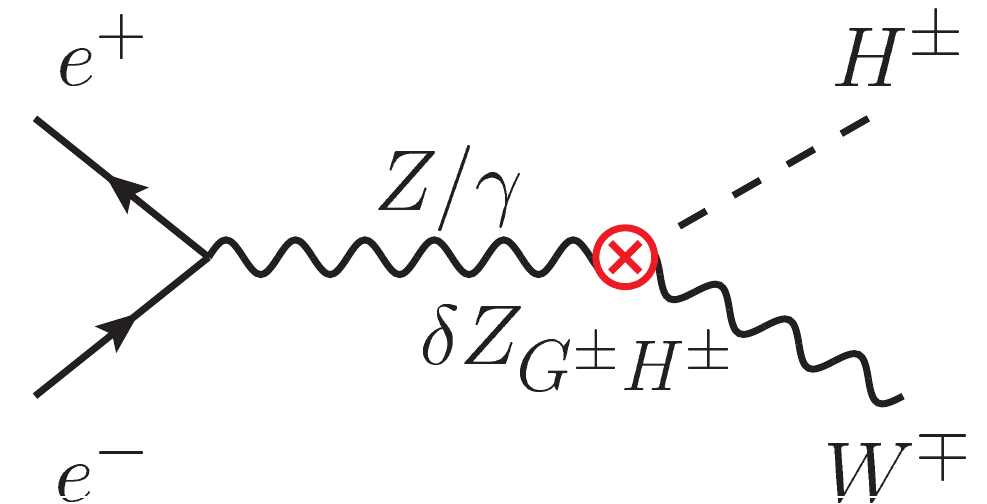}
    \caption{Representative Feynman diagrams for $e^{+}e^{-} \rightarrow H^{\pm}W^{\mp}$ at NLO in QCD.}
    \label{fig:feynman_nlo}
\end{figure*}
\begin{figure*}
    \centering
    \includegraphics[scale=0.4]{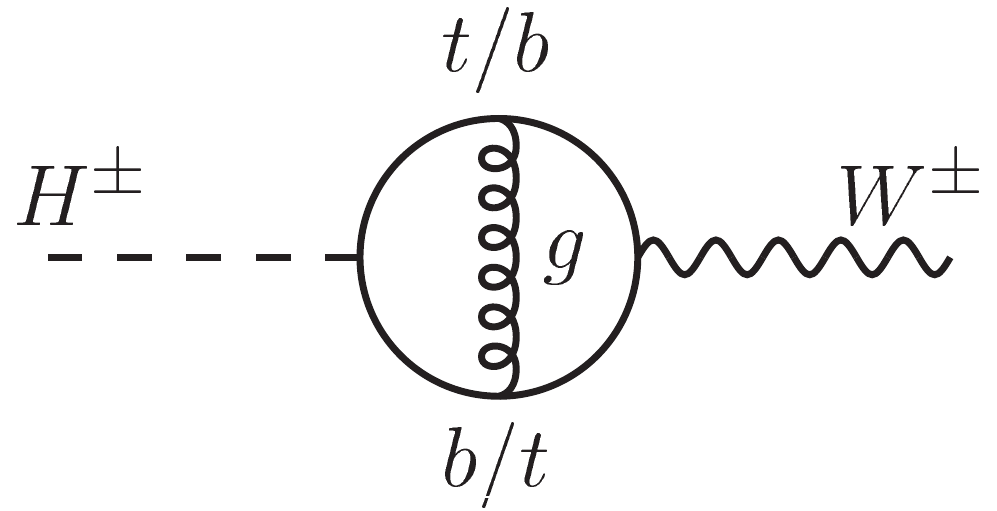}
    \includegraphics[scale=0.4]{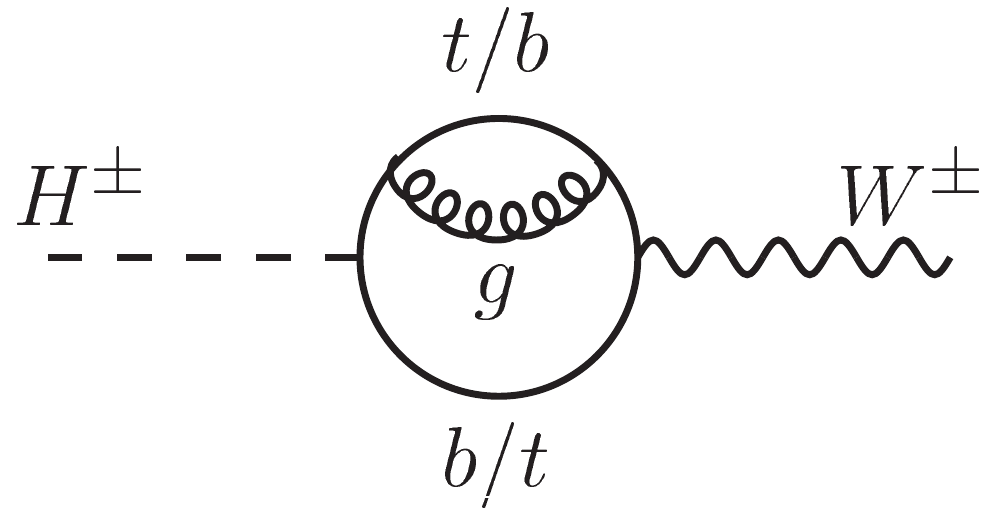}
    \includegraphics[scale=0.4]{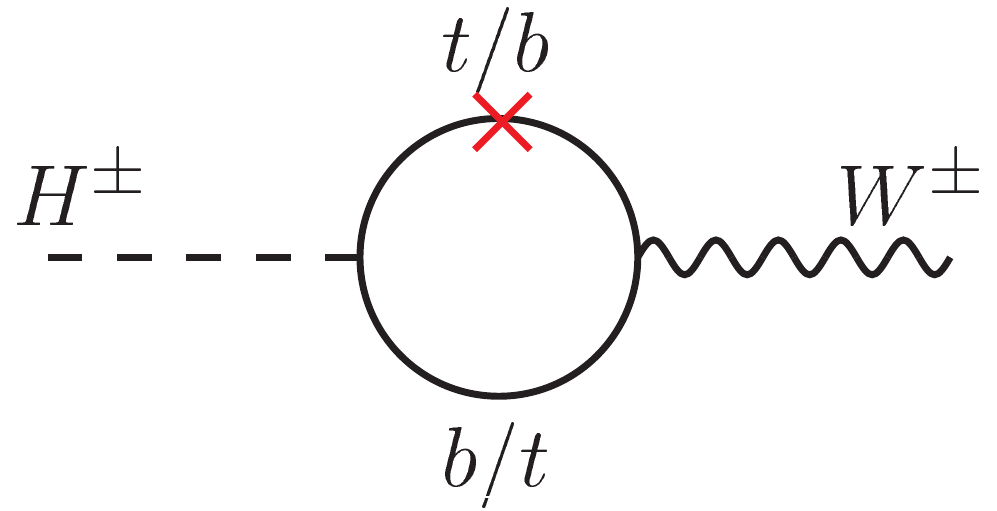}
    \caption{Representative Feynman diagrams contributing to $W^{\pm}-H^{\pm}$ transition at $\mathcal{O}(\alpha\alpha_{s})$.}
    \label{fig:mixing}
\end{figure*}

There are $24$ two-loop and counterterm diagrams for $e^{+}e^{-} \rightarrow H^{\pm}W^{\mp}$ at the QCD NLO. Some representative ones of them are depicted in Fig.\ref{fig:feynman_nlo}. At the QCD NLO, all the two-loop Feynman diagrams are generated from the LO quark triangle loop diagram (i.e., the first diagram in Fig.\ref{fig:feynman_lo}). The cross in the quark loop diagrams represents the renormalization constant of quark mass at $\mathcal{O}(\alpha_{s})$, while the circle cross displayed in the last counterterm diagram represents the renormalization constant $\delta Z_{G^{\pm}H^{\pm}}$ at $\mathcal{O}(\alpha\alpha_{s})$. The quark mass renormalization constant used in NLO QCD calculation is given by \cite{Bernreuther:2004ih}
\begin{equation}
    \delta_{m_{q}} = -m_{q}\frac{\alpha_{s}}{2\pi}C(\epsilon)\left(\frac{\mu^{2}}{m_{q}^{2}}\right)^{\epsilon}\frac{C_{F}}{2}\frac{(3-2\epsilon)}{\epsilon(1-2\epsilon)},
\end{equation}
where $C(\epsilon) = (4\pi)^{\epsilon}\Gamma(1+\epsilon)$ and $C_{F}=\dfrac{4}{3}$. The lowest order for $e^{+}e^{-} \rightarrow H^{\pm}W^{\mp}$ is one-loop order, therefore, the renormalization should be dealt with carefully in NLO QCD calculation. As shown in Figs.\ref{fig:feynman_lo} and \ref{fig:feynman_nlo}, the wave-function renormalization constant $\delta Z_{G^{\pm}H^{\pm}}$ is involved in both NLO and LO amplitudes. Since the self-energy $\sum^{W^{\pm}H^{\pm}}(m_{H^{\pm}}^2)$ is nonzero at $\mathcal{O}(\alpha\alpha_{s})$, i.e. $\delta Z_{G^{\pm}H^{\pm}}$ is nonzero at $\mathcal{O}(\alpha\alpha_{s})$, the contribution from the last diagram in Fig.\ref{fig:feynman_nlo} should be included in NLO QCD calculation. The typical Feynman diagrams for $H^{\pm}-W^{\pm}$ transition at $\mathcal{O}(\alpha\alpha_{s})$ are shown in Fig.\ref{fig:mixing}. After taking into account all the contributions at $\mathcal{O}(\alpha\alpha_{s})$ in $D$-dimensional spacetime, both $\dfrac{1}{\epsilon^2}$ and $\dfrac{1}{\epsilon}$ singularities are all canceled.

\section{\label{sec:results}Numerical results and discussion}
\par
Besides the input parameters for the Higgs sector of the THDM specified in benchmark scenario in Eq.(\ref{eq:benchmark}), the following SM input parameters are adopted in our numerical calculation \cite{Tanabashi:2018oca}:
\begin{equation}
    \begin{aligned}
         & G_{F} = 1.1663787\times 10^{-5} ~{\rm GeV^{-2}}, \quad \alpha_{s}(m_Z) = 0.118,                \\
         & m_{t} = 173 ~{\rm GeV}, \quad m_{b} = 4.78 ~{\rm GeV},       \\
         & m_{W} = 80.379 ~{\rm GeV}, \quad m_{Z} = 91.1876 ~{\rm GeV},
    \end{aligned}
\end{equation}
where $G_{F}$ is the Fermi constant. The fine structure constant $\alpha$ is fixed by
\begin{equation}
    \alpha=\frac{\sqrt{2}G_{F}}{\pi}\frac{m_W^2(m_Z^2-m_W^2)}{m_Z^2}.
\end{equation}

\subsection{LO}
\par
\begin{figure*}
    \centering
    \includegraphics[scale=0.29]{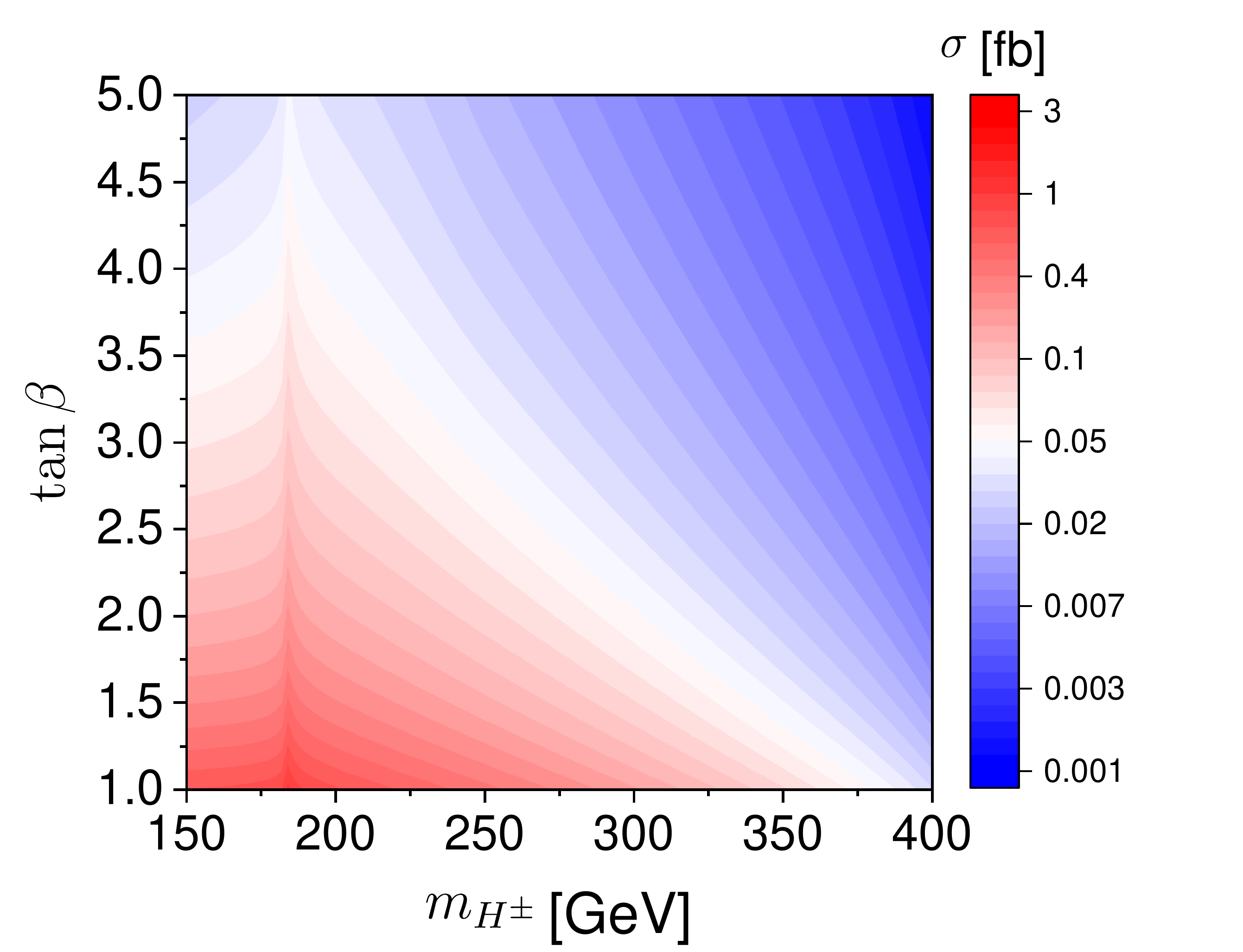}
    \includegraphics[scale=0.29]{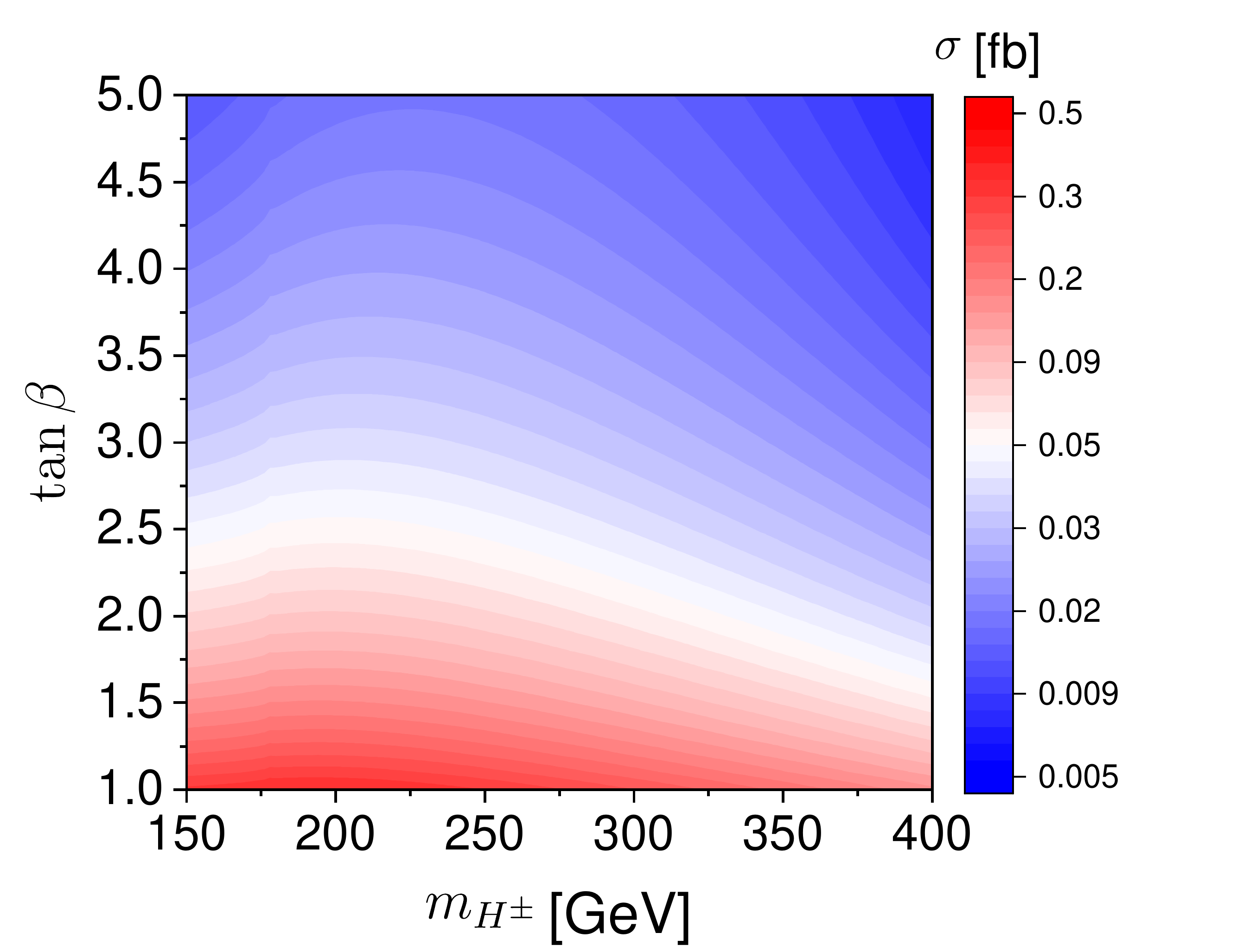}
    \caption{
    Contours of LO cross section for $e^{+}e^{-} \rightarrow H^{\pm}W^{\mp}$ at $\sqrt{s} = 500~ {\rm GeV}$ (left) and $1000~ {\rm GeV}$ (right) on the $m_{H^{\pm}}-\tan\beta$ plane.}
    \label{fig:scan_mhp_tb}
\end{figure*}

In Fig.\ref{fig:scan_mhp_tb}, we display the LO cross section for $e^{+}e^{-} \rightarrow H^{\pm}W^{\mp}$ as a function of $m_{H^{\pm}}$ and $\tan\beta$ in the benchmark scenario in Eq.(\ref{eq:benchmark}) at $\sqrt{s} = 500~ {\rm GeV}$ (left) and $1000~ {\rm GeV}$ (right), respectively. From the left plot, we can see clearly that the LO cross section for $H^{\pm}W^{\mp}$ production at a $500~ {\rm GeV}$ $e^+e^-$ collider peaks at $m_{H^{\pm}} \simeq 184~ {\rm GeV}$ due to the resonance effect of loop integrals. The cross section is sensitive to the mass of charged Higgs boson, it can exceed $3~ {\rm fb}$ in the vicinity of $m_{H^{\pm}} \simeq 184~ {\rm GeV}$ at small $\tan\beta$. In the region of $m_{H^{\pm}}<180~ {\rm GeV}$, the cross section increases slowly as the increment of $m_{H^{\pm}}$, while it drops rapidly when $m_{H^{\pm}} > 184~ {\rm GeV}$. As the increment of $\tan\beta$ from $1$ to $5$, the cross section decreases consistently due to the decline of the $H^{+}\bar{t}b$ Yukawa coupling strength in the low $\tan\beta$ region. Comparing the two plots in Fig.\ref{fig:scan_mhp_tb}, we can see that the cross section at $\sqrt{s} = 1000~ {\rm GeV}$ is much smaller than that at $\sqrt{s} = 500~ {\rm GeV}$ because of the $s$-channel suppression. As the increasing of the $e^+e^-$ colliding energy from $500$ to $1000~ {\rm GeV}$, the peak position of the cross section as a function of $m_{H^{\pm}}$ moves towards high $m_{H^{\pm}}$ and the $m_{H^{\pm}}$ dependence of the cross section is reduced significantly. Moreover, the production cross section at $\sqrt{s}=1000~ {\rm GeV}$ also decreases quickly as the increment of $\tan\beta$ in the plotted $\tan\beta$ region.

\par
\begin{figure*}
    \centering
    \includegraphics[scale=0.29]{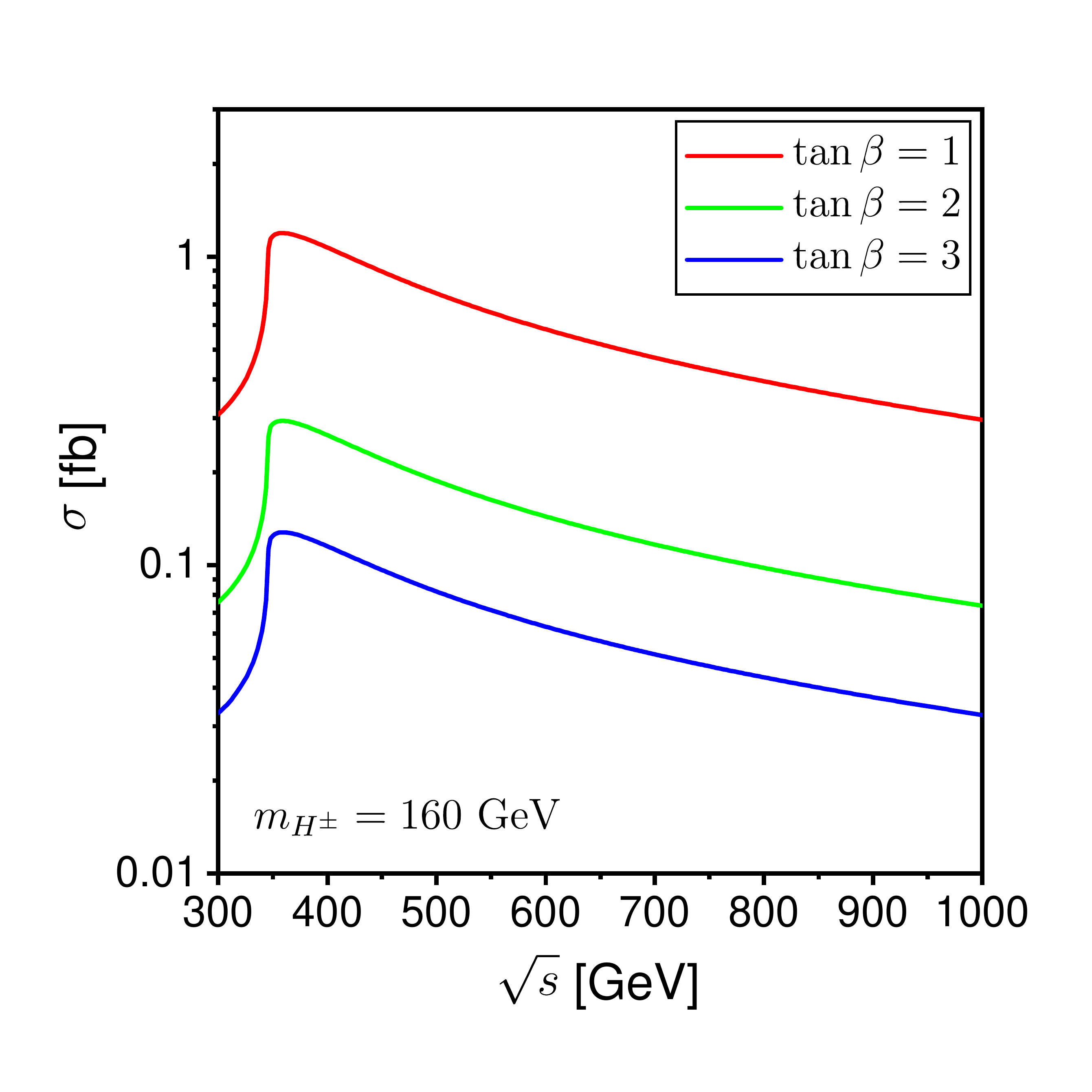}
    \includegraphics[scale=0.29]{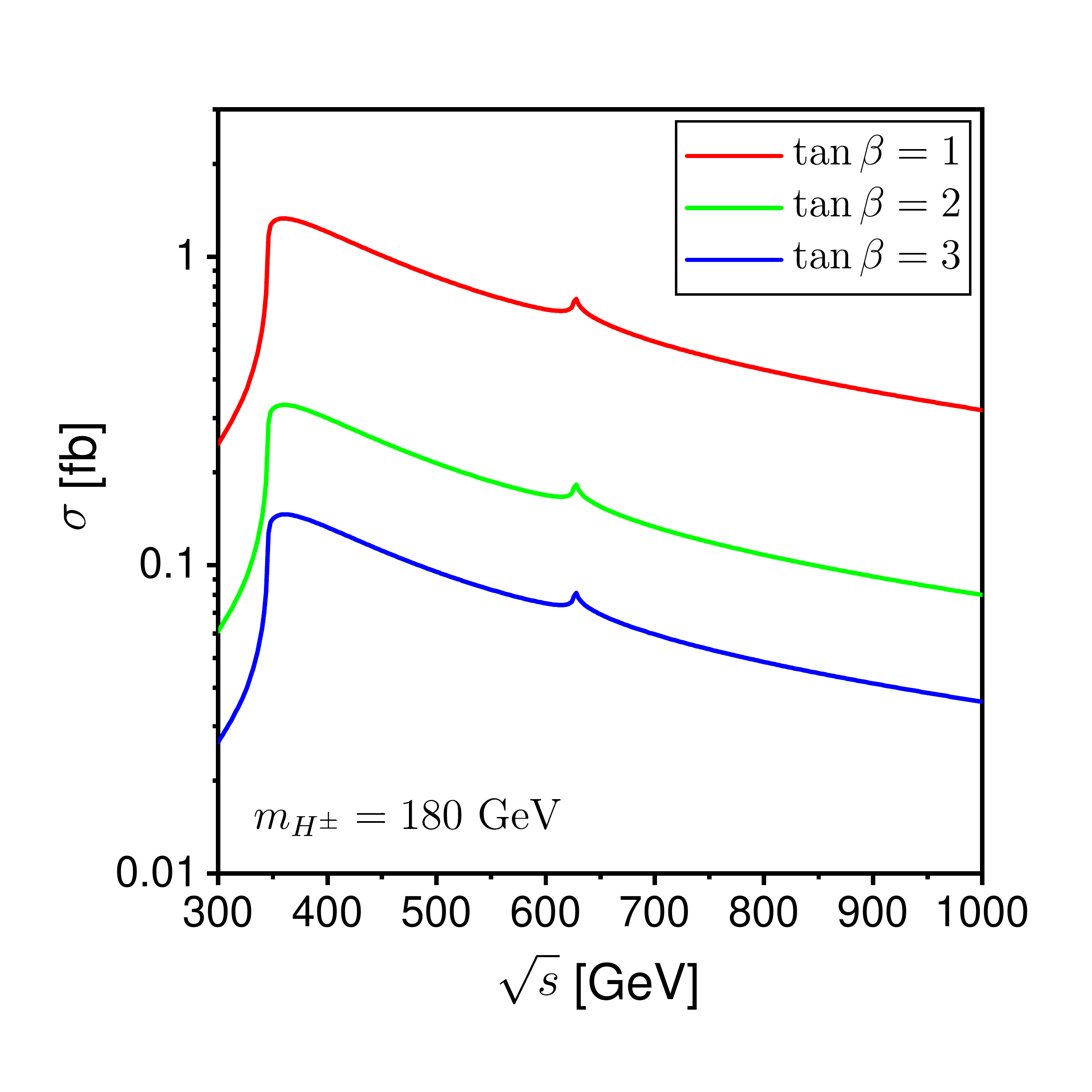}\\
    \includegraphics[scale=0.29]{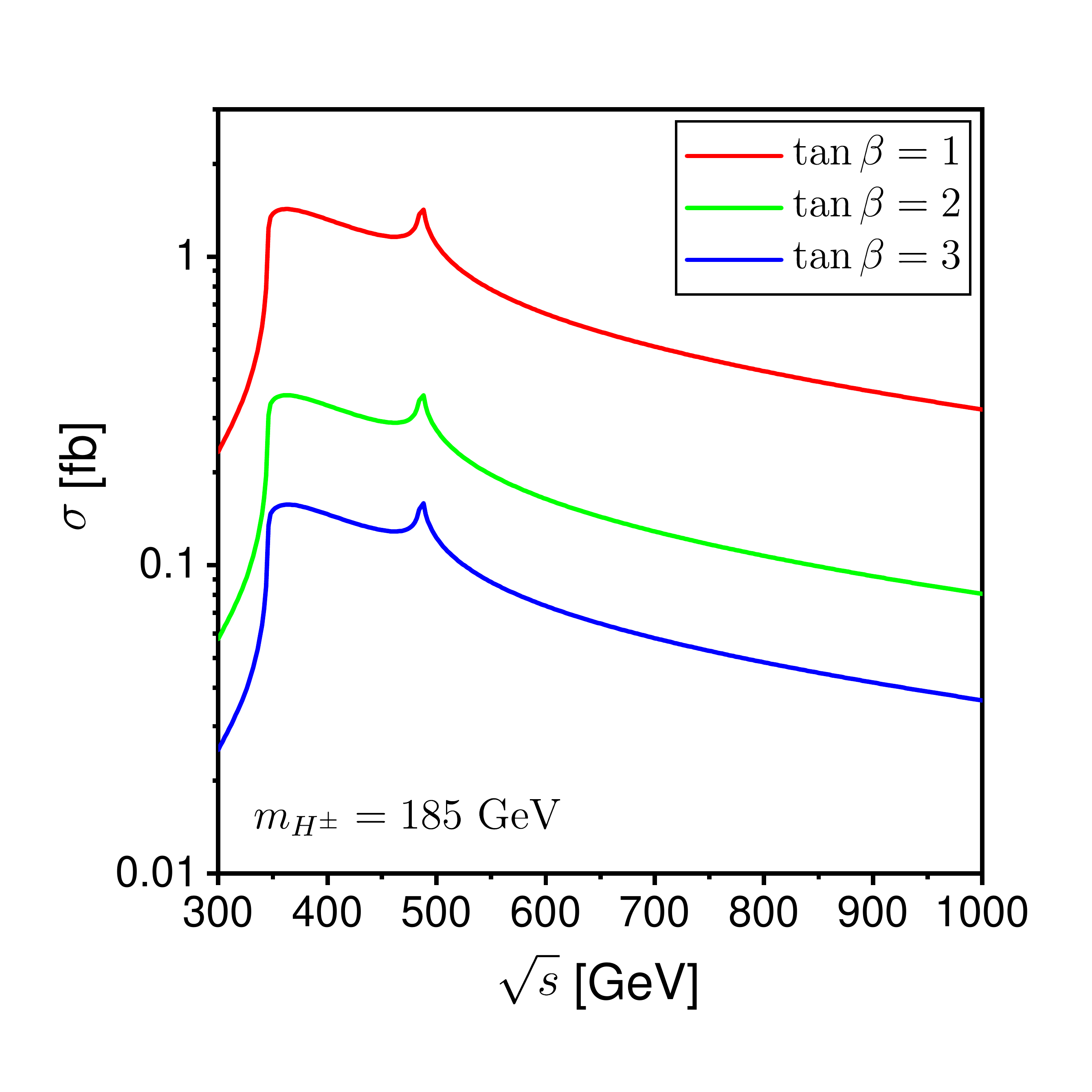}
    \includegraphics[scale=0.29]{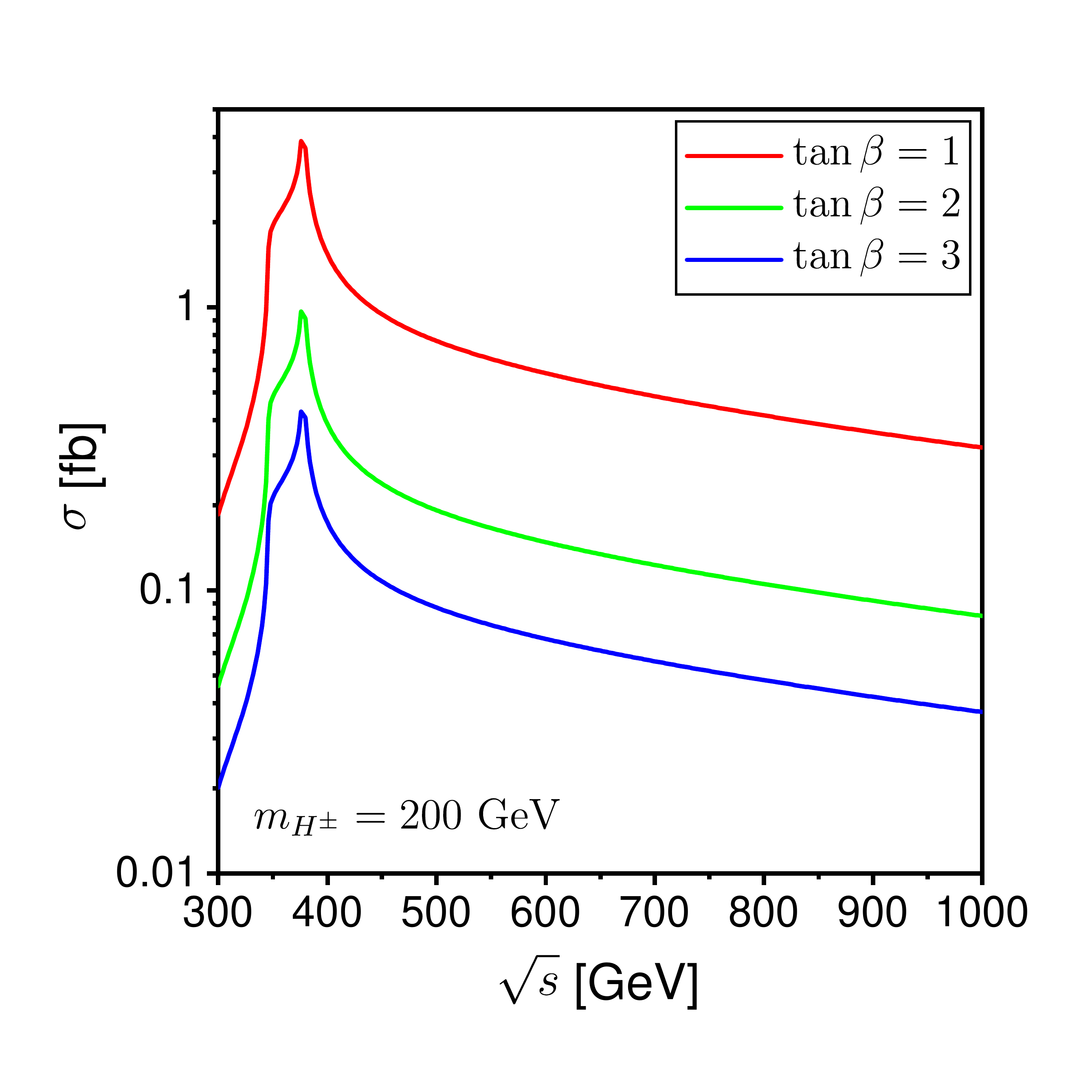}
    \caption{LO cross section for $e^{+}e^{-} \rightarrow H^{\pm}W^{\mp}$ as a function of $e^+e^-$ colliding energy for some typical values of $m_{H^{\pm}}$ and $\tan\beta$.}
    \label{fig:scan_sqrts_LO}
\end{figure*}

In Fig.\ref{fig:scan_sqrts_LO}, we present the dependence of the LO cross section for $e^+e^- \rightarrow H^{\pm}W^{\mp}$ on the $e^+e^-$ colliding energy for some typical values of $m_{H^{\pm}}$ and $\tan\beta$. As shown in this figure, the behaviors of the production cross section as a function of the colliding energy at different values of $\tan\beta$ are quite similar. For $m_{H^{\pm}}=160~ {\rm GeV}$, the cross section increases sharply in the range of $\sqrt{s} < 360~ {\rm GeV}$, reaches its maximum at $\sqrt{s} \simeq 360~ {\rm GeV}$, and then decreases slowly as the increment of $\sqrt{s}$. The existence of the peak at $\sqrt{s} \simeq 360~ {\rm GeV}$ can be attributed to the competition between the phase-space enlargement and the $s$-channel suppression as the increasing of $\sqrt{s}$. The maximum value of the cross section can exceed $1~ {\rm fb}$ for $\tan\beta = 1$ and decreases to about $0.3~ {\rm fb}$ and $0.1~ {\rm fb}$ for $\tan\beta = 2$ and $\tan\beta = 3$, respectively. Comparing the upper two plots of Fig.\ref{fig:scan_sqrts_LO}, we can see that the $\sqrt{s}$ dependence of the cross section for $m_{H^{\pm}} = 180~ {\rm GeV}$ is very close to that for $m_{H^{\pm}} = 160~ {\rm GeV}$, but there is a small peak at $\sqrt{s} \simeq 630~ {\rm GeV}$ for $m_{H^{\pm}} = 180~ {\rm GeV}$. Such resonance induced by loop integrals only occurs above the threshold of $H^{+} \rightarrow t \bar{b}$, i.e., $m_{H^{\pm}} > m_t + m_b$. As the increment of $m_{H^{\pm}}$, this resonance effect becomes more considerable and the peak position moves towards low $\sqrt{s}$. As shown in the bottom-left plot of Fig.\ref{fig:scan_sqrts_LO}, the resonance peak for $m_{H^{\pm}} = 185~ {\rm GeV}$ is located at $\sqrt{s} \simeq 490~ {\rm GeV}$, and is more distinct compared to that for $m_{H^{\pm}} = 180~ {\rm GeV}$. As to the $\sqrt{s}$ dependence of the $H^{\pm}W^{\mp}$ production cross section for $m_{H^{\pm}} = 200~ {\rm GeV}$ shown in the bottom-right plot of Fig.\ref{fig:scan_sqrts_LO}, it looks quite different from those for $m_{H^{\pm}}= 160,\, 180$ and $185~ {\rm GeV}$. There is a sharp peak at $\sqrt{s} \simeq 390~ {\rm GeV}$ for each value of $\tan\beta \in \{1, 2, 3\}$ which was also mentioned in previous works \cite{Zhu:1999ke,Kanemura:1999tg,Heinemeyer:2016wey}. This peak is a consequence of competition among the phase-space enlargement, $s$-channel suppression and the resonance induced by loop integrals. We can see that the peak cross section can reach about $4~ {\rm fb}$ for $\tan\beta = 1$, and will decrease to around $1~ {\rm fb}$ and $0.4~ {\rm fb}$ when $\tan\beta$ increases to $2$ and $3$, respectively. In the region of $\sqrt{s} < 350~ {\rm GeV}$, the cross section increases quickly as the increment of $\sqrt{s}$ due to the enlargement of phase space, while in the region $\sqrt{s} > 450~ {\rm GeV}$, it is close to the result of $s$-channel suppression, especially at high $\sqrt{s}$.

\subsection{NLO QCD}
\par
In this subsection, we calculate the $e^+e^- \rightarrow H^{\pm}W^{\mp}$ process at the QCD NLO, and discuss the dependence of the integrated cross section on the $e^+e^-$ colliding energy and the charged Higgs mass as well as the angular distribution of the final-state charged Higgs boson.

\par
\begin{figure}
    \centering
    \includegraphics[scale=0.35]{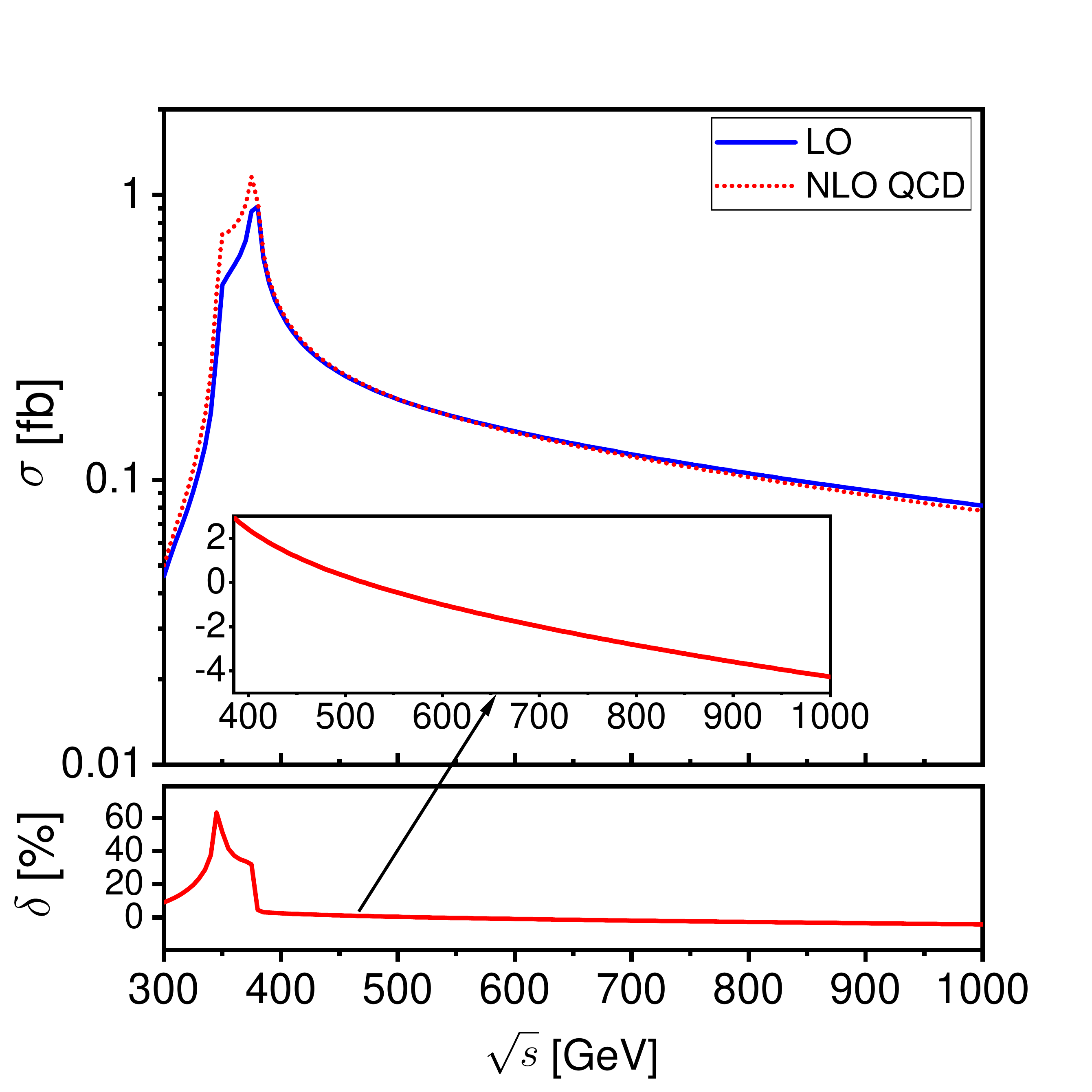}
    \caption{LO, NLO QCD corrected integrated cross sections and the corresponding QCD relative correction for $e^+e^- \rightarrow H^{\pm}W^{\mp}$ as functions of $e^+e^-$ colliding energy for $m_{H^{\pm}} = 200~ {\rm GeV}$ and $\tan\beta = 2$.}
    \label{fig:scan_sqrts_NLO}
\end{figure}

The LO, NLO QCD corrected integrated cross sections and the corresponding QCD relative correction for $e^+e^- \rightarrow H^{\pm}W^{\mp}$ as functions of the $e^+e^-$ colliding energy $\sqrt{s}$ are depicted in Fig.\ref{fig:scan_sqrts_NLO}, where $m_{H^{\pm}}=200~ {\rm GeV}$ and $\tan\beta = 2$. As shown in the upper panel of this figure, the NLO QCD corrected integrated cross section peaks at $\sqrt{s} \simeq 375~ {\rm GeV}$, it increases sharply when $\sqrt{s} < 375~ {\rm GeV}$ and decreases approximately linearly in the region of $\sqrt{s} > 500~ {\rm GeV}$ as the increment of $\sqrt{s}$. From the lower panel of Fig.\ref{fig:scan_sqrts_NLO}, we can see that the QCD relative correction increases rapidly from about $9\%$ to above $60\%$ as the increment of $\sqrt{s}$ from $300$ to $345~ {\rm GeV}$ and then decreases back to about $3\%$ as $\sqrt{s}$ increases to $385~ {\rm GeV}$. The variation of QCD relative correction with $\sqrt{s}$ in the region of $\sqrt{s} > 385~ {\rm GeV}$ is also plotted in the inset in the upper panel of Fig.\ref{fig:scan_sqrts_NLO} for clarity. It clearly shows that the QCD relative correction decreases approximately linearly from about $3\%$ to around $-4\%$ as the increment of $\sqrt{s}$ from $385$ to $1000~ {\rm GeV}$. In Table \ref{tab:scan_sqrts_NLO} we list the LO, NLO QCD corrected cross sections and the corresponding QCD relative corrections at some specific colliding energies. At $\sqrt{s} = 340~ {\rm GeV}$ which can be reached by both the International Linear Collider (ILC) \cite{Behnke:2013xla} and the Future Circular Electron-Positron Collider (FCC-ee) \cite{Abada:2019zxq}, the cross section is about $0.235~ {\rm fb}$ at the QCD NLO. At the ILC with $\sqrt{s} = 500$ and $1000~ {\rm GeV}$, the NLO QCD corrected cross sections reach about $0.193$ and $0.0778~ {\rm fb}$, respectively, and the corresponding relative corrections are $0.26\%$ and $-4.27\%$.
\begin{table*}
    \begin{ruledtabular}
        \begin{tabular}{ccccccccccc}
            $\sqrt{s}$ [GeV]             & $300$     & $320$     & $340$    & $400$    & $500$    & $600$    & $700$    & $800$    & $900$     & $1000$    \\
            \hline
            $\sigma_{\textrm{LO}}$ [fb]  & $0.04592$ & $0.07868$ & $0.1712$ & $0.3870$ & $0.1920$ & $0.1481$ & $0.1230$ & $0.1054$ & $0.09196$ & $0.08126$ \\
            $\sigma_{\textrm{NLO}}$ [fb] & $0.05004$ & $0.09163$ & $0.2353$ & $0.3963$ & $0.1925$ & $0.1466$ & $0.1206$ & $0.1024$ & $0.08865$ & $0.07779$ \\
            $\delta$ [\%]                & $8.97$    & $16.4$    & $37.4$   & $2.40$   & $0.260$  & $-1.01$  & $-1.95$  & $-2.85$  & $-3.60$   & $-4.27$   \\
        \end{tabular}
    \end{ruledtabular}
    \caption{LO, NLO QCD corrected cross sections and the corresponding QCD relative corrections for $e^+e^- \rightarrow H^{\pm}W^{\mp}$ at some specific colliding energies. ($m_{H^{\pm}}=200~ {\rm GeV}$ and $\tan\beta = 2$)}
    \label{tab:scan_sqrts_NLO}
\end{table*}

\par
\begin{figure}
    \centering
    \includegraphics[scale=0.35]{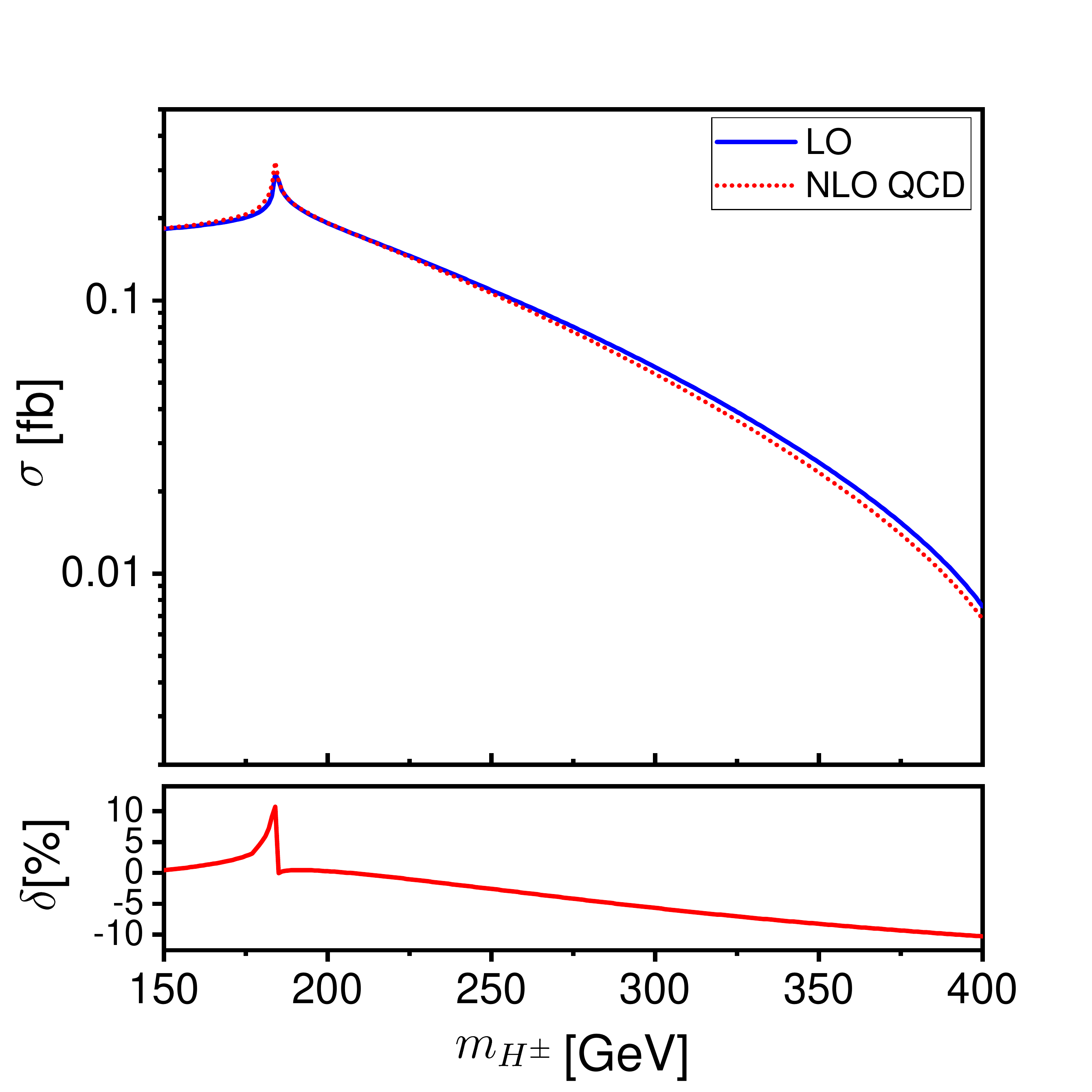}
    \caption{LO, NLO QCD corrected cross sections and the corresponding QCD relative correction for $H^{\pm}W^{\mp}$ associated production at a $\sqrt{s} = 500~ {\rm GeV}$ $e^+e^-$ collider as functions of charged Higgs mass for $\tan\beta=2$.}
    \label{fig:scan_mhp_NLO}
\end{figure}

In order to study the $m_{H^{\pm}}$ dependence of the QCD correction, we plot the NLO QCD corrected cross section, as well as the LO cross section, and the QCD relative correction as functions of $m_{H^{\pm}}$ in Fig.\ref{fig:scan_mhp_NLO} with $\tan\beta=2$ and $\sqrt{s}=500~ {\rm GeV}$. The numerical results for some typical values of $m_{H^{\pm}}$ are also given in Table \ref{tab:scan_mhp_NLO}. From Fig.\ref{fig:scan_mhp_NLO}, we can see that the LO and NLO QCD corrected cross sections increase from about $0.18~ {\rm fb}$ to around $0.29$ and $0.32~ {\rm fb}$, respectively, as $m_{H^{\pm}}$ increases from $150$ to $184~ {\rm GeV}$, and drop to less than $0.01~ {\rm fb}$ when $m_{H^{\pm}}=400~ {\rm GeV}$. Similarly, there is also a notable spike at $m_{H^{\pm}} \simeq 184~ {\rm GeV}$ for QCD relative correction, as shown in the lower panel of Fig.\ref{fig:scan_mhp_NLO}. The QCD relative correction is less than $0.5\%$ and thus can be neglected when $m_{H^{\pm}} = 150~ {\rm GeV}$, while it is expected to increase to about $11\%$ as $m_{H^{\pm}}$ increases to $184~ {\rm GeV}$. When $m_{H^{\pm}} > 185~ {\rm GeV}$, the QCD relative correction decreases slowly down to about $-10\%$ as $m_{H^{\pm}}$ increases to $400~ {\rm GeV}$.
\begin{table*}
    \begin{ruledtabular}
        \begin{tabular}{ccccccccccc}
            $m_{H^{\pm}}$ [GeV]          & $150$    & $160$    & $170$    & $180$    & $190$    & $200$    & $250$    & $300$     & $350$     & $400$      \\
            \hline
            $\sigma_{\textrm{LO}}$ [fb]  & $0.1828$ & $0.1876$ & $0.1951$ & $0.2140$ & $0.2230$ & $0.1920$ & $0.1090$ & $0.05699$ & $0.02553$ & $0.007595$ \\
            $\sigma_{\textrm{NLO}}$ [fb] & $0.1836$ & $0.1896$ & $0.1990$ & $0.2249$ & $0.2240$ & $0.1925$ & $0.1062$ & $0.05376$ & $0.02343$ & $0.006817$  \\
            $\delta$ [\%]                & $0.438$   & $1.07$   & $2.00$   & $5.09$   & $0.448$ & $0.260$  & $-2.57$  & $-5.67$   & $-8.22$   & $-10.24$   \\
        \end{tabular}
    \end{ruledtabular}
    \caption{LO, NLO QCD corrected cross sections and the corresponding QCD relative corrections for $H^{\pm}W^{\mp}$ production at a $\sqrt{s} = 500~ {\rm GeV}$ $e^+e^-$ collider for some typical values of $m_{H^{\pm}}$. ($\tan\beta=2$)}
    \label{tab:scan_mhp_NLO}
\end{table*}

\par
\begin{figure*}
    \centering
    \includegraphics[scale=0.29]{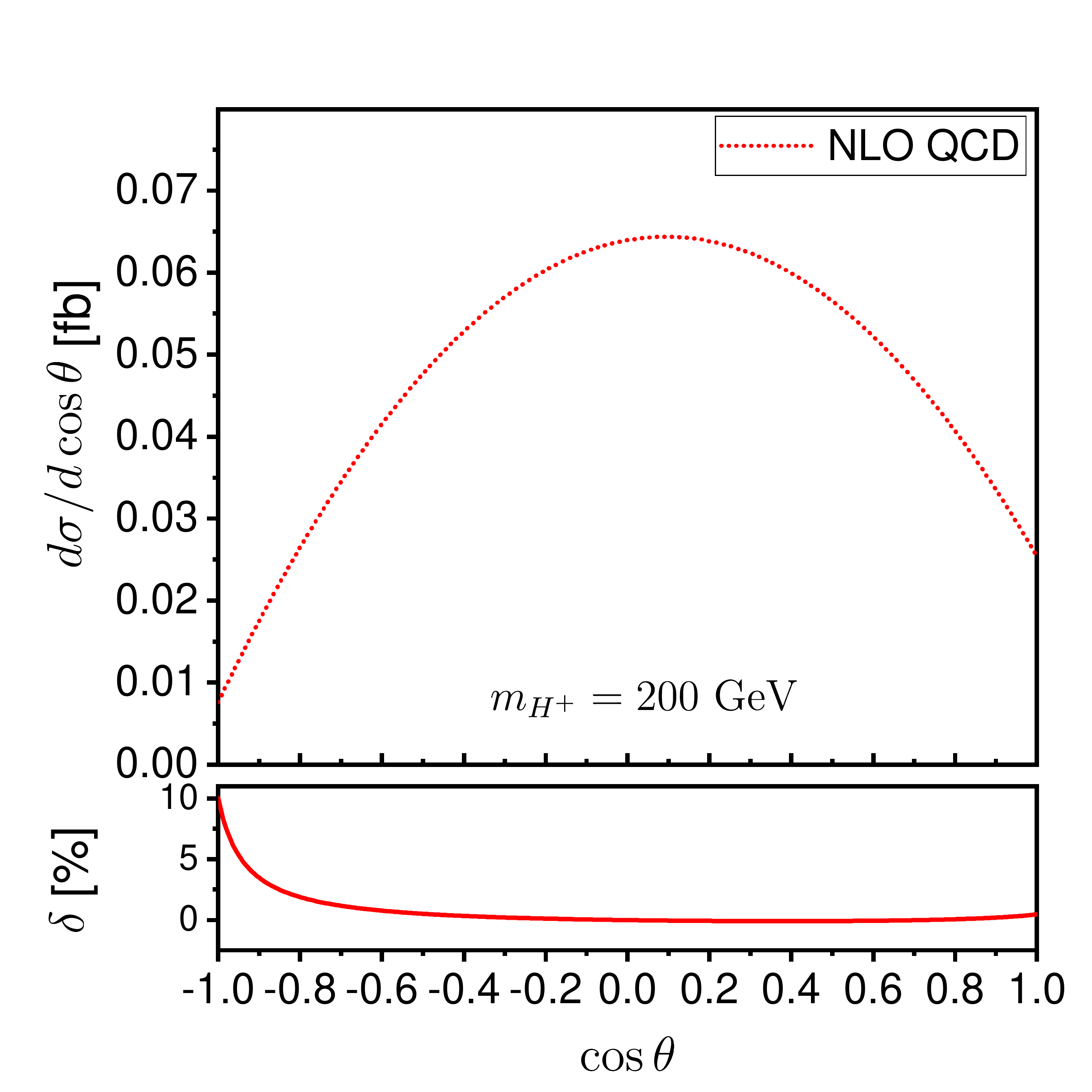}
    \includegraphics[scale=0.29]{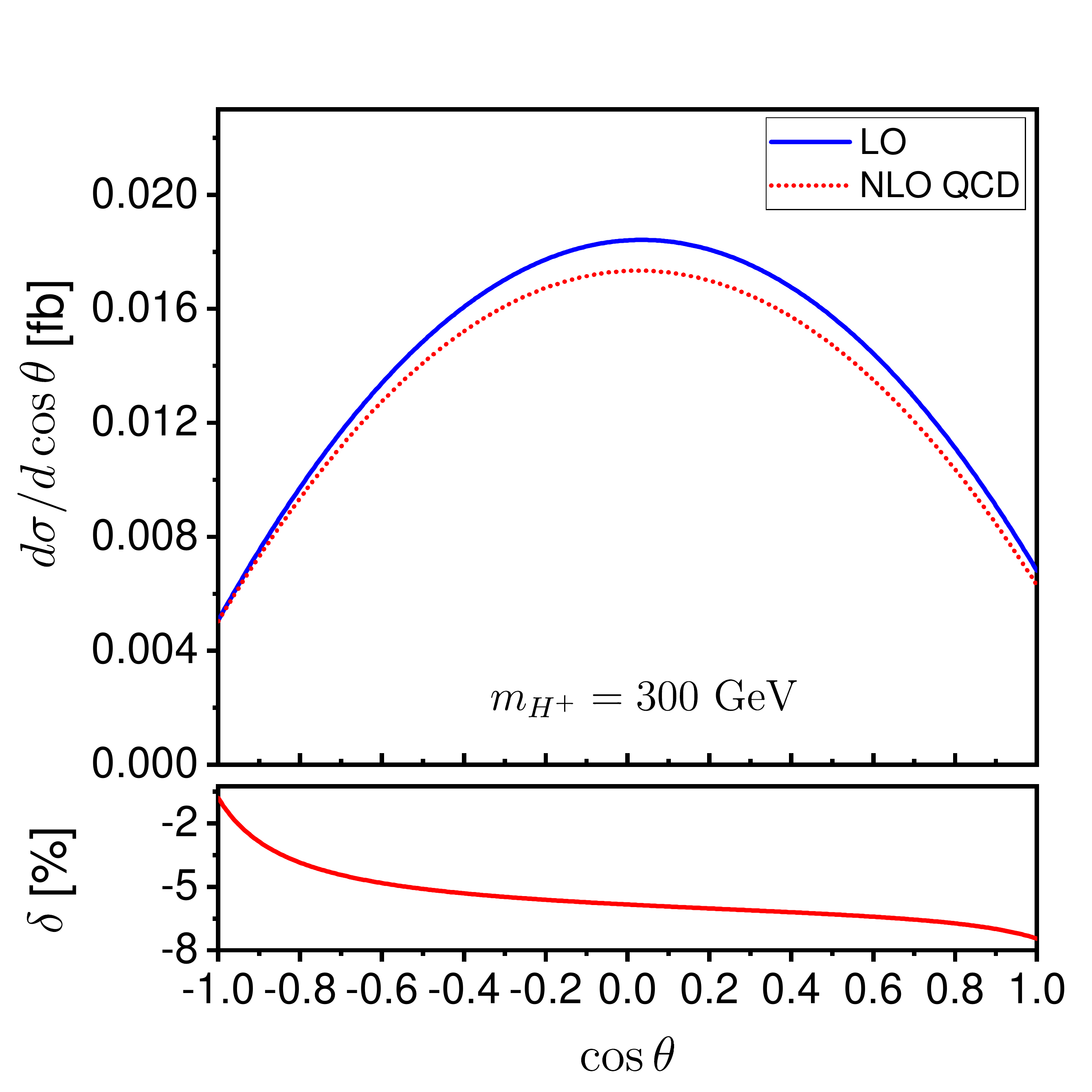}
    \caption{Angular distributions of charged Higgs boson for $H^+W^-$ associated production at a $500~ {\rm GeV}$ $e^+e^-$ collider for $\tan\beta = 2$ and $m_{H^{\pm}}=200$ (left) and $300~ {\rm GeV}$ (right).}
    \label{fig:diff_mhp}
\end{figure*}

The LO, NLO QCD corrected angular distributions of the final-state charged Higgs boson and the corresponding QCD relative corrections for $H^+W^-$ associated production at a $500~ {\rm GeV}$ $e^+e^-$ collider for $\tan\beta = 2$ and $m_{H^{\pm}} = 200$ and $300~ {\rm GeV}$ are depicted in Fig.\ref{fig:diff_mhp}, where $\theta$ denotes the scattering angle of $H^+$ with respect to the electron beam direction. Due to the $\mathcal{CP}$ conservation, the distribution of the scattering angle of $H^-$ with respect to the positron beam direction for $e^+e^- \rightarrow H^-W^+$ is the same as the angular distribution of $H^+$ for $e^+e^- \rightarrow H^+W^-$. From this figure, we can see that the charged Higgs boson is mostly produced in transverse direction for both $m_{H^{\pm}} = 200$ and $300~ {\rm GeV}$. For $m_{H^{+}}=200~ {\rm GeV}$, the QCD relative correction decreases rapidly from $10\%$ to nearly $0\%$ as the increment of $\cos\theta$ from $-1$ to $-0.5$, and is steady at around $0\%$ in the region of $-0.5 < \cos\theta < 1$. It implies that the NLO QCD correction can be neglected in most of the phase space region except when $\theta \rightarrow \pi$. For $m_{H^{\pm}} = 300~ {\rm GeV}$, the LO differential cross section is suppressed by the NLO QCD correction in the whole phase space region. The corresponding QCD relative correction decreases from about $-0.8\%$ to $-7.5\%$ as $\cos\theta$ increases from $-1$ to $1$. This QCD correction should be taken into consideration for precision study of the $H^{\pm}W^{\mp}$ production at lepton colliders.

\section{\label{sec:summary}Summary}
Searching for exotic Higgs boson and studying its properties are important tasks at future lepton colliders. In this work, we study in detail the $H^{\pm}W^{\mp}$ associated production at future electron-positron colliders within the framework of the Type-I THDM. We calculate the $e^+e^- \rightarrow H^{\pm}W^{\mp}$ process at the LO, and investigate the dependence of the production cross section on the THDM parameters ($m_{H^{\pm}}$ and $\tan\beta$) and the $e^+e^-$ colliding energy. The numerical results show that the cross section is very sensitive to the charge Higgs mass in the vicinity of $m_{H^{\pm}} \simeq 184~ {\rm GeV}$ at a $500~ {\rm GeV}$ $e^+e^-$ collider, and decreases consistently with the increment of $\tan\beta$ in the low $\tan\beta$ region. The existence of a peak in the colliding energy distribution of the cross section is explained by the resonance effect induced by loop integrals. This resonance occurs only above the threshold of $H^+ \rightarrow t \bar{b}$, and the peak position moves towards low colliding energy as the increment of $m_{H^{\pm}}$. We also calculate the two-loop NLO QCD corrections to $e^+e^- \rightarrow H^{\pm}W^{\mp}$, and provide some numerical results for the NLO QCD corrected integrated cross section and the angular distribution of the final-state charged Higgs boson. For $\sqrt{s} = 500~ {\rm GeV}$ and $\tan\beta = 2$, the QCD relative correction varies smoothly in the range of $[-10\%,\, 3\%]$ as the increment of $m_{H^{\pm}}$ from $150$ to $400~ {\rm GeV}$, except in the vicinity of $m_{H^{\pm}} \simeq 184~ {\rm GeV}$. The QCD relative correction is sensitive to the charged Higgs mass and strongly depends on the final-state phase space. For $m_{H^{\pm}} = 300~ {\rm GeV}$ and $\tan\beta = 2$, the QCD relative correction to the $H^+W^-$ production at a $500~ {\rm GeV}$ $e^+e^-$ collider increases from about $-7.5\%$ to $-0.8\%$ as the scattering angle of $H^+$ increases from $0$ to $\pi$. Compared to hadron colliders, the measurement precision of Higgs associated production at future high-energy electron-positron colliders is much higher. The expected experimental error of Higgs production in association with a weak gauge boson at high-energy electron-positron colliders is less than $1\%$ through the recoil mass of the associated vector boson. For example, the measurement precision of $HZ$ production at $\sqrt{s} = 240~ {\rm GeV}$ FCC-ee and $\sqrt{s} = 250~ {\rm GeV}$ CEPC can reach about $0.4\%$ and $0.7\%$, respectively \cite{Peskin:2012we,Gomez-Ceballos:2013zzn,Ruan:2014xxa}. Due to the high $W$-tagging efficiency at $e^+e^-$ colliders, the measurement precision of $H^{\pm}W^{\mp}$ production at a high-energy $e^+e^-$ collider can be also less than $1\%$. Thus, the two-loop QCD correction should be taken into consideration in precision study of the $H^{\pm}W^{\mp}$ associated production at future lepton colliders.

\section{\label{sec:ac}Acknowledgements}
This work is supported in part by the National Natural Science Foundation of China (Grants No. 11775211 and No. 11535002) and the CAS Center for Excellence in Particle Physics (CCEPP).

\bibliography{eehw}

\end{document}